\documentclass[10pt]{article}
\usepackage[a4paper, total={6in, 8in}]{geometry}
\usepackage[utf8]{inputenc}
\usepackage[T1]{fontenc}    
\usepackage{graphicx}
\usepackage[table]{xcolor}
\usepackage{todonotes}
\usepackage{subcaption}
\usepackage{makecell}
\usepackage{tabularx}

\usepackage{booktabs} 
\usepackage{xparse}   

\NewDocumentCommand{\rot}{O{45} O{1em} m}{\makebox[#2][l]{\rotatebox{#1}{#3}}}%

\newcolumntype{b}{X}
\newcolumntype{s}{>{\hsize=.25\hsize}X}
\newcolumntype{t}{>{\hsize=.075\hsize}X}

\usepackage{natbib}
\title{Is the Contralateral Delay Activity (CDA) a robust neural correlate for Visual Working Memory (VWM) tasks? \\A reproducibility study.}
\date{} 

\begin{document}

\author{
  \textbf{Yannick Roy} \\
  Faubert Lab \\
  Universit\'e de Montr\'eal\\
  Montr\'eal, Canada \\
  \texttt{yannick.roy@umontreal.ca}
  \and
  \textbf{Jocelyn Faubert} \\
  Faubert Lab \\
  Universit\'e de Montr\'eal\\
  Montr\'eal, Canada \\
}

\maketitle

\section*{Abstract}
Visual working memory (VWM) allows us to actively store, update and manipulate visual information surrounding us. While the underlying neural mechanisms of VWM remain unclear, contralateral delay activity (CDA), a sustained negativity over the hemisphere contralateral to the positions of visual items to be remembered, is often used to study VWM. To investigate if the CDA is a robust neural correlate for VWM tasks, we reproduced eight CDA-related studies with a publicly accessible EEG dataset. We used the raw EEG data from these eight studies and analysed all of them with the same basic pipeline to extract CDA. We were able to reproduce the results from all the studies and show that with a basic automated EEG pipeline we can extract a clear CDA signal. We share insights from the trends observed across the studies and raise some questions about the CDA decay and the CDA during the recall phase, which surprisingly, none of the eight studies did address. Finally, we also provide reproducibility recommendations based on our experience and challenges in reproducing these studies.



\section{Introduction} 
\label{sec:introduction}

Visual working memory (VWM) allows us to actively store, update and manipulate visual information surrounding us. Acting as a mental buffer for visual information, VWM has been an active area of research for several decades. While behavioural studies have shown that working memory (WM), of which VWM is a subset, has a limited capacity of only a few items, usually ranging between 3 to 5 (\cite{cowan2001magical, bays2008dynamic}), the underlying neural mechanisms remain vague. One ongoing challenge in the field is dissociating WM from attention (\cite{cowan1998attention, olivers2011different, awh2006interactions, oberauer2002access}). There is no clear consensus yet as how separated or intertwined these two mechanisms really are.
 
Given the crucial role of VWM in our everyday life, much prior work has tried to understand the neural correlates underlying its functioning via electroencephalogam (EEG). One such VWM neural measurement being studied is the contralateral delay activity (CDA) (\cite{vogel2004neural, mccollough2007electrophysiological}). CDA is a sustained negativity over the hemisphere contralateral to the positions of the items to be remembered. A prevailing view of CDA is that it is modulated by the number of items held in WM reaching a plateau at around three or four items. The CDA has been shown to be linked with the number of items held in WM (\cite{vogel2004neural, unsworth2015working, luria2016contralateral}).

In order to study CDA, the most common task is a change detection task. The sequence of such task typically looks something like the following: the participant is cued with an array of items, varying in numbers but usually between one and eight, balanced on both sides of the visual field. The items then disappear, forcing the participant to hold relevant information in mind and after a short period of time, usually one or two seconds, the participant is quizzed on the visual representation held in WM. For example, a new array of items can be presented and the participant is asked if there has been a change versus the initial items. The participant will answer with a keyboard or a mouse and the results will be logged with different levels of granularity depending on what the study is interested in. There are many variants of such tasks.

Many well-known event-related potentials (ERPs) such as the P300 and N270, benefit from a large volume of studies and replications and are well understood (\cite{polich1995cognitive, rossion2000n170, fazel2012p300, luck2014introduction}). CDA, however, doesn't benefit from such volume of evidence yet. Our primary goal with this reproducibility study was to answer the following question: \textit{Is the Contralateral Delay Activity (CDA) a robust and consistent neural correlate for Visual Working Memory (VWM) tasks?}. We wanted to know if CDA is a consistent measure across different tasks, subjects and EEG recording devices, or if it requires a lot of manual cleaning and handcrafting of the data to obtain it. To investigate the robustness of the CDA EEG signal, we looked for CDA-related EEG datasets available online and tried to reproduce their results using a basic independent automated pipeline with no human intervention on the data to extract the CDA.

Lastly, before we dive in, we should define the terms \textit{reproducibility} and \textit{replicability} given that researchers from different fields, and even from within the same field, often use them interchangeably. Here we will use the same definitions as in the \textit{Recommendations to Funding Agencies for Supporting Reproducible Research} by the American Statistical Association (\cite{broman2017recommendations}):

\textbf{Reproducibility}: \textit{A study is reproducible if you can take the original data and the computer code used to analyze the data and reproduce all of the numerical findings from the study. This may initially sound like a trivial task but experience has shown that it’s not always easy to achieve this seemingly minimal standard.}
        
\textbf{Replicability}: \textit{This is the act of repeating an entire study, independently of the original investigator without the use of original data (but generally using the same methods).}

Simply put, one can replicate a study or an effect (outcome of a study) but reproduce results (data analyses). A third term, \textit{repeatability}, is also used although less often, referring to the same group repeating the same experiment with the same analysis and obtaining the same results.

This current work focuses on reproducibility and not on replicability nor repeatability as we did not collect new data but only used publicly available datasets. This reproducibility study also intersects with a review given that we extract common trends among studies and explore them from a computational point of view. Note that we initially reproduced the result either with the original code or with a re-written version of it before using the same pipeline for all studies to assess the robustness.

Other fields such as artificial intelligence (AI) have benefited tremendously from good reproducibility frameworks, standards and overall culture. The accelerated pace of innovation and breakthroughs in AI were mainly enabled by accessibility of both data and code. 
These best practices of sharing both data and code in a reproducible manner aren't, unfortunately, the default behavior in psycho- and neuro-related fields. Here, we share how a few groups of researchers have made their data and code available and we hope to inspire others to follow that path.


\section{Method}

After looking through CDA-related literature, searching for available datasets and asking (i.e. emailing) a few researchers in the field if they were aware of open access CDA-related EEG datasets, we ended up with eight recent CDA-related studies published between 2018 and 2020 with different task paradigms. We do not claim that the list of studies included in this review represents an exhaustive list of CDA papers with available EEG datasets, however, we believe that these eight datasets represent a good sample of the CDA literature as they used different VWM tasks exploring things such as different set sizes (from 1 to 6 targets), adding new targets after the initial array, retro-cueing, adding targets bilaterally, adding interruptions, halving the targets to create more items to track, and all that using different shapes and colors as stimuli across studies. In this section, we detail the studies we've reproduced and provide the resulting CDA signal figure, which matches the CDA figure from the original study. It is important to note that some of the reproduced studies were also looking at different neural correlates, however, for simplicity and readability we focus only on the CDA relevant part of the study. In our analysis, we used all the available data, but did not include any data files that were clearly marked as not being used (for various reasons). All the studies were reproduced using MNE-Python (\cite{gramfort2013meg}) despite the original code of most studies being in MATLAB. All figures in this section were generate using the following pipeline: (1) load raw data, (2) rereference and downsample based on the original study's preprocessing steps, (3) filter the data between 1 and 30 Hz, (4) epoch the data, (5) automated removal of bad trials (i.e. epochs) and interpolation of bad channels via \emph{autoreject} (\cite{jas2017autoreject}), (6) obtain each subject's CDA for each condition by subtracting the averaged ipsilateral electrodes to the averaged contralateral electrodes (i.e. contra minus ipsi), (7) average the CDA for each condition across subjects.


Table \ref{table_datasets} shows the high-level information of the datasets to provide the reader an idea about the number of subjects and trials for each study.

\begin{table} [!htb]
\small
\begin{tabularx}{\textwidth}{tsb}
\hline
\textbf{Year} & \textbf{First Author} & \textbf{Title} \\
\hline\hline
2020 & Tobias Feldmann‐Wüstefeld & Neural measures of working memory in a bilateral change detection task \\
\hline
2020 & Nicole Hakim & Perturbing neural representations of working memory with task-irrelevant interruption \\ 
\hline
2019 & Mario Villena-Gonzalez & Data from brain activity during visual working memory replicates the correlation between contralateral delay activity and memory capacity \\ 
\hline
2019 & Haley Balaban & Neural evidence for an object-based pointer system underlying working memory \\ 
\hline
2019 & Eren Gunseli & EEG dynamics reveal a dissociation between storage and selective attention within working memory \\ 
\hline
2019 & Nicole Hakim & Dissecting the Neural Focus of Attention Reveals Distinct Processes for Spatial Attention and Object-Based Storage in Visual Working Memory \\ 
\hline
2018 & Kristen Adam & Contralateral delay activity tracks fluctuations in working memory performance \\ 
\hline
2018 & Tobias Feldmann‐Wüstefeld & Contralateral Delay Activity Indexes Working Memory Storage, not the Current Focus of Spatial Attention \\ 
\hline
\end{tabularx}
\caption{Reproduced Studies}
\label{table_papers}
\end{table}

\begin{table} [!ht]
\begin{tabular}{|c|c|c|c|c|}
\hline
\textbf{Dataset} & \textbf{Task} & \textbf{Subjects} & \textbf{Trials} & \textbf{Target(s)} \\
\hline\hline
FW2020 & Change Detection Task & 21 & 1560 & 2,4,6 \\
\hline
H2020 & Change Detection Task & 22/20 & 2400/1920 & 4 \\
\hline
B2019 & (Bilateral) Change Detection Task & 16/16/16 & 840/660/840 & 2,4 \\
\hline
H2019 & Change Detection Task & 97 & 1600 & 2,4 \\
\hline
G2019 & Orientation Retro-Cued Task & 30 & 500 & 1, 3* \\
\hline
VG2019 & Change Detection Task & 23 & 96 & 1,2,4 \\
\hline
FW2018 & (Sequential) Change Detection Task & 23/20 & 960/960 & 1,2,3,4 \\
\hline
A2018 & Lateralized Whole-Report Task & 31/48 & 650/540 & 1,3,6 \\
\hline
\end{tabular}
\caption{Datasets - Details}
\label{table_datasets}
\end{table}


\subsection{Feldmann-Wüstefeld et al., 2020}
\begin{figure} [!htb]
    \centering
    \includegraphics[width=0.5\textwidth]{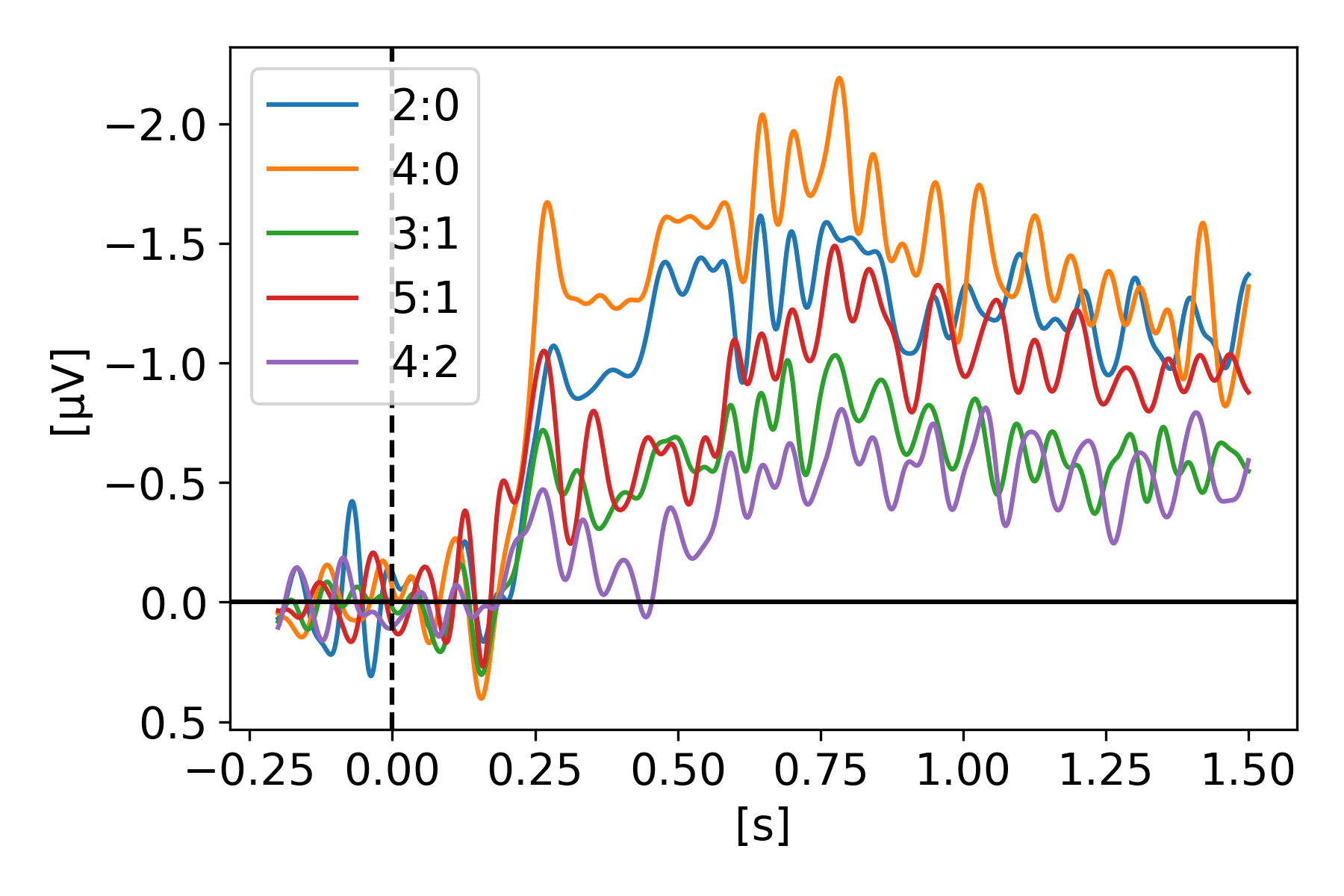}
    \caption{Reproduced results from Feldmann-Wüstefeld et al., 2020}
    \label{Feldmann-Wustefeld-2020-reprod}
\end{figure}

In \textit{Neural measures of working memory in a bilateral change detection task} (\cite{feldmann2021neural}), Feldmann-Wüstefeld and colleagues used a novel change detection task in which both the CDA and the negative slow wave (NSW) can be measured at the same time. They presented memory items bilaterally with different set sizes in both hemifields inducing an imbalance or “net load” as they called it. Their results showed that the NSW increased with set size, whereas the CDA increased with net load. There were three different set sizes participants had to remember: two, four or six targets. In Figure \ref{Feldmann-Wustefeld-2020-reprod}, we can see the five combinations of targets they used (\textit{2:0, 3:1, 4:2, 4:0, 5:1}). With their nomenclature \textit{2:0} means 2 targets in one hemifield and 0 target in the other hemifield for a net load of 2. \textit{4:2} represents 6 targets total, 4 in one hemifield and 2 in the other for a net load of 2. As we can see, the highest CDA is obtained with \textit{4:0} for a net load of 4. Interestingly, the \textit{5:1} condition shows a higher CDA than 3:1 and 4:2 however lower than 2:0 indicating that having targets in both hemifields reduces the overall CDA amplitude. On the graph, t=0s is when the memory display appeared on the screen with the targets, then they stayed visible for 500ms after which the participant had to remember the targets for 1 second before the probe display appeared for the participant to provide their answer as to confirm if the item in the probe display was indeed part of the targets or not. 

\subsection{Hakim et al., 2020}
\begin{figure} [!htb]
    \centering
    \includegraphics[width=0.5\textwidth]{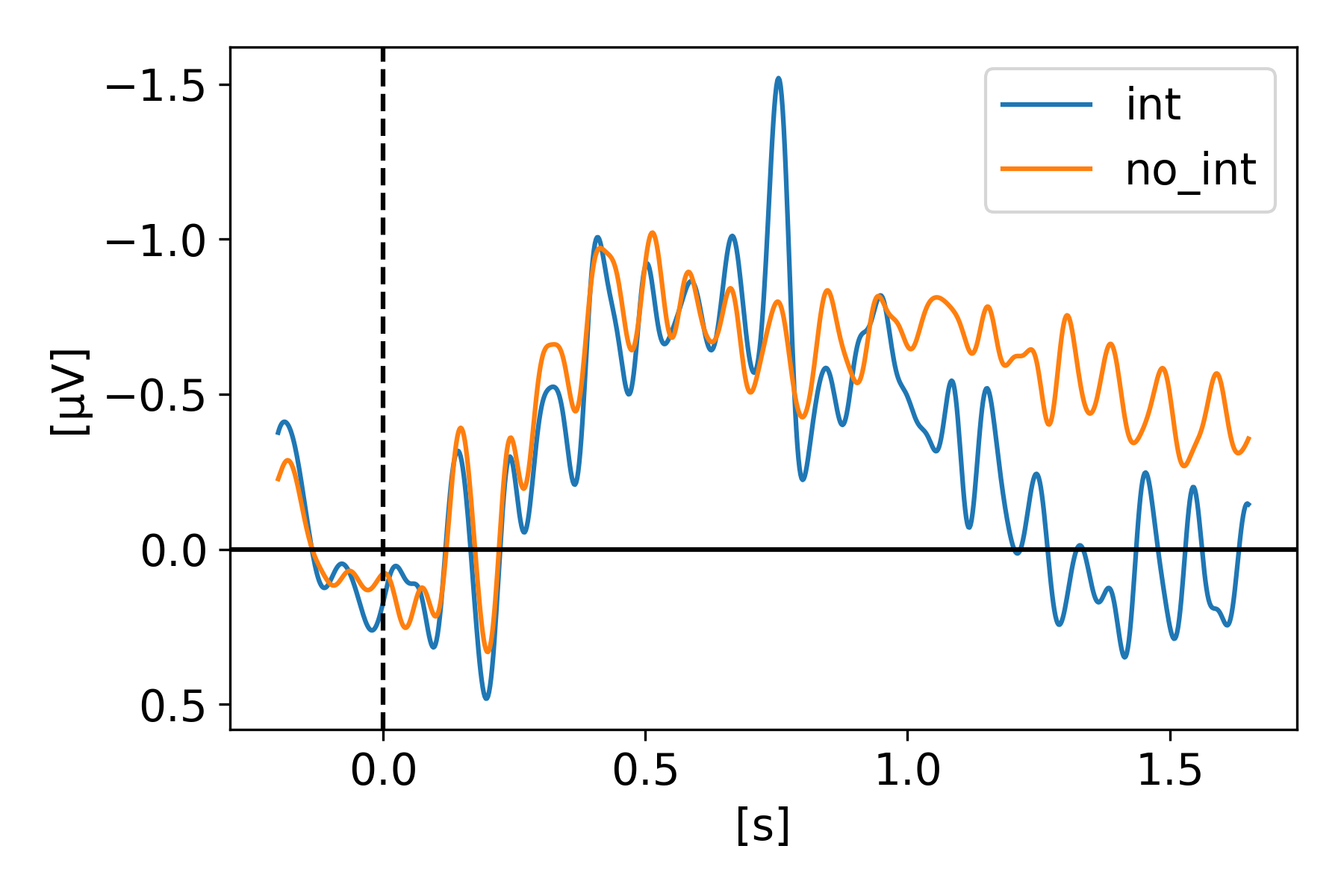}
    \caption{Reproduced results from Hakim et al., 2020 - Experiment 1}
    \label{Hakim-2020-Exp1-reprod}
\end{figure}

In \textit{Perturbing neural representations of working memory with task-irrelevant interruption} (\cite{hakim2020perturbing}), Hakim and colleagues investigated the impact of task-irrelevant interruptions on neural representations of working memory across two experiments looking at both the CDA and lateralized alpha power. What they found is that after interruption, the CDA amplitude momentarily sustained but was gone by the end of the trial. On the other hand, lateralized alpha power, which has been used as an effective tool for discerning which visual hemifield is being attended, was immediately influenced by the interrupters but recovered by the end of the trial. Hakim and colleagues suggested that dissociable neural processes contribute to the maintenance of working memory information and that brief irrelevant onsets disrupt two distinct online aspects of working memory and also that task-irrelevant interruption could motivate the transfer of information from active to passive storage, explaining the reason why the CDA drops significantly after interruptions, even on trials with good performance (i.e. the participant remembered the targets correctly). In their 2nd experiment, they go further and test if the expectation of interruption changes the CDA. The full 1.65s epoch is displayed but not the answer part which took place after 1.65s (see \ref{Hakim-2020-Exp1-reprod}). The CDA was obtained by using only PO7 and PO8 electrodes. 

\subsection{Balaban et al., 2019}
\begin{figure} [!htb]
\centering
\begin{subfigure}{.49\textwidth}
    \centering
    \includegraphics[width=.95\linewidth]{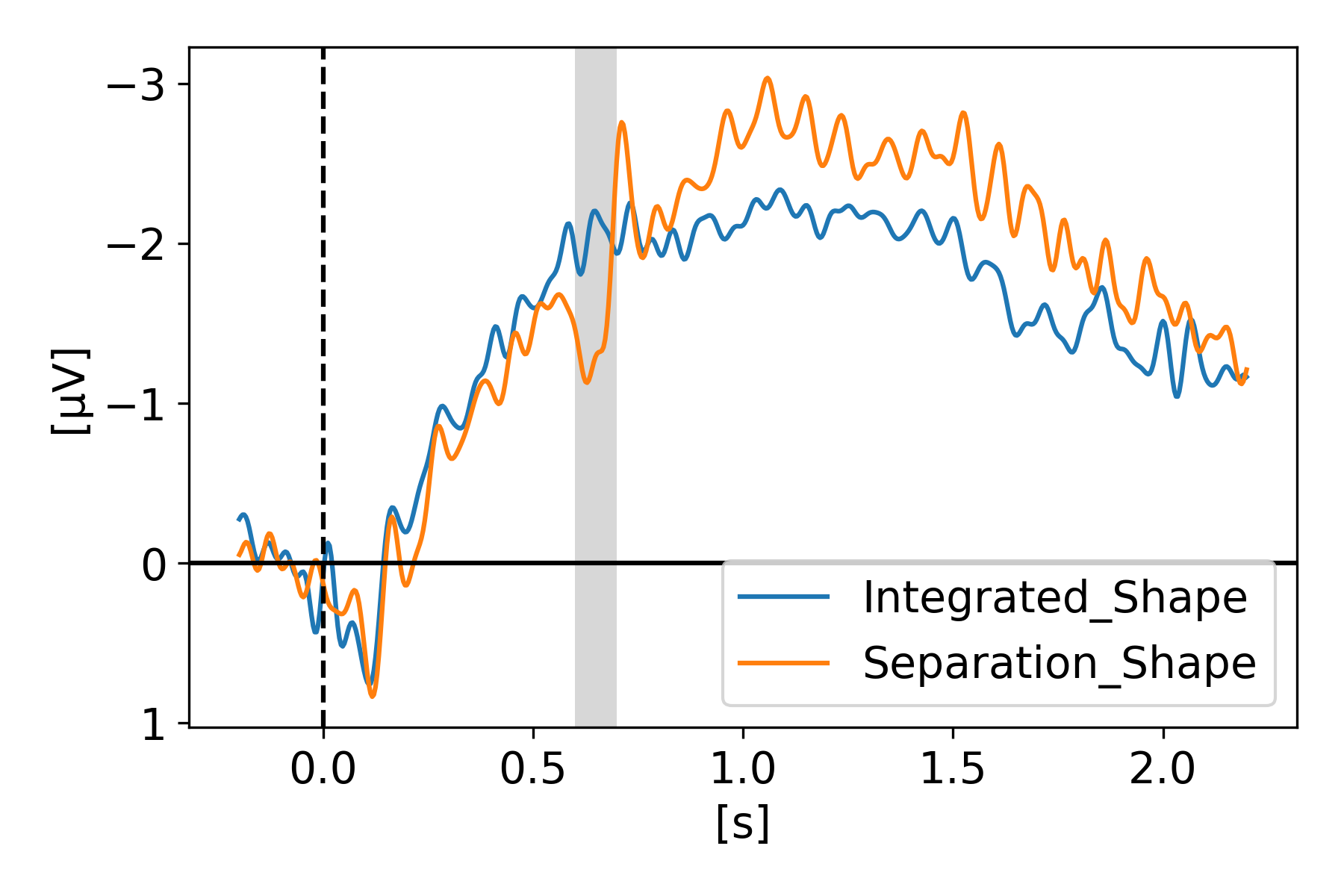}
\end{subfigure}
\begin{subfigure}{.49\textwidth}
    \centering
    \includegraphics[width=.95\linewidth]{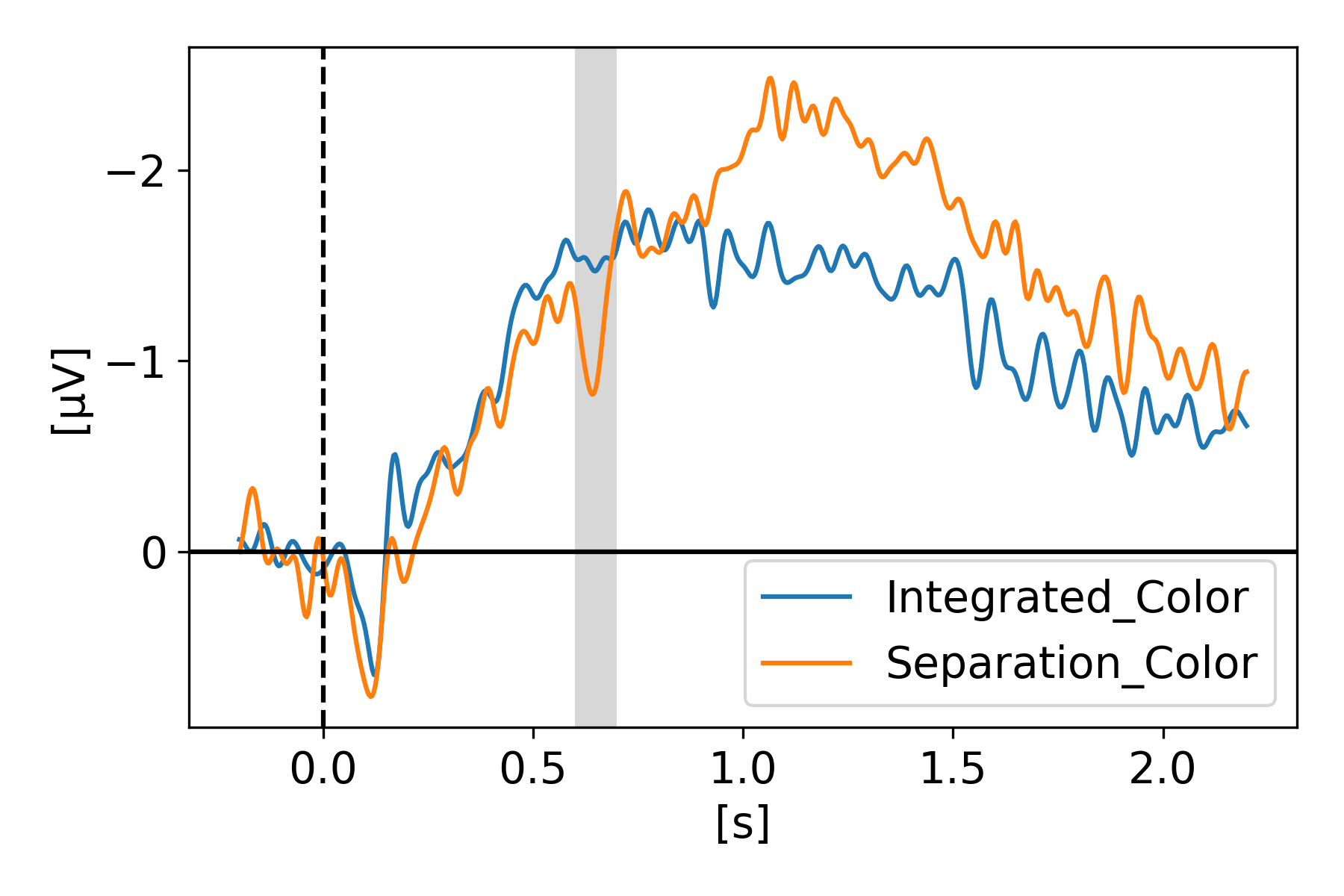}
\end{subfigure}
\caption{Reproduced results from Balaban et al., 2019 - Experiment 1}
\label{Balaban-2019-Exp1-reprod}
\end{figure}

In \textit{Neural evidence for an object-based pointer system underlying working memory} (\cite{balaban2019neural}), Balaban and colleagues argued that to update our representation of the environment, our VWM depends on a pointer system such that each representation is stably and uniquely mapped to a specific stimulus. Therefore, without these pointers, our VWM representations are inaccessible. Via three experiments, they examined whether the pointers are allocated in an object-based, featural, or spatial manner. Their results showed that the separation of an object in two halves invalidated the pointers. It happened in a shape task, where the separation changed both the objects and the task-relevant features, but also in a color task, where the separation destroyed the objects while leaving the task-relevant features intact. They suggested that objects, and not task-relevant features, underlie the pointer system. Two of their three experiments are displayed in Figure \ref{Balaban-2019-Exp1-reprod} and Figure \ref{Balaban-2019-Exp2-reprod}, while the third experiment is available in the supplementary material to reduce the length of this manuscript and enhance readability.

For Experiment 1, t=0s is when the memory display appeared on the screen with the moving objects which either separated in halves after 400ms or either continued moving as a whole for another 600ms after which they stopped moving for 300ms and disappeared. After 900ms of retention with an empty screen displaying only a fixation cross in the middle, the participant had to give an answer as if the objects being displayed are the same or different than the initial ones. In one condition the shapes were the relevant features (i.e. have some of the shapes changed?) and in a second condition the colors were the relevant features (i.e. have some of the colors changed?). On the graph, the cognitive impact of the separation that happened for half the trials (at 400ms) is clearly visible in the CDA signal around 200ms as highlighted by the grayed region. The full 2.2s epoch is displayed on Figure \ref{Balaban-2019-Exp1-reprod} but not the answer part which took place after 2.2s. The CDA was obtained by using only PO7-PO8, P7-P8 and PO3-PO4 electrode pairs. 

For Experiment 2 (Figure \ref{Balaban-2019-Exp2-reprod}), the epochs were shorter in time and only the colors were the relevant features. The targets were all moving squares that could either separate after 400ms or continue moving as a whole. Some trials had two targets and some trials had four targets. In both experiments we see that after the separation, the CDA amplitude increases, as the number of targets to track has now increased.

\begin{figure} [!htb]
    \centering
    \includegraphics[width=0.5\textwidth]{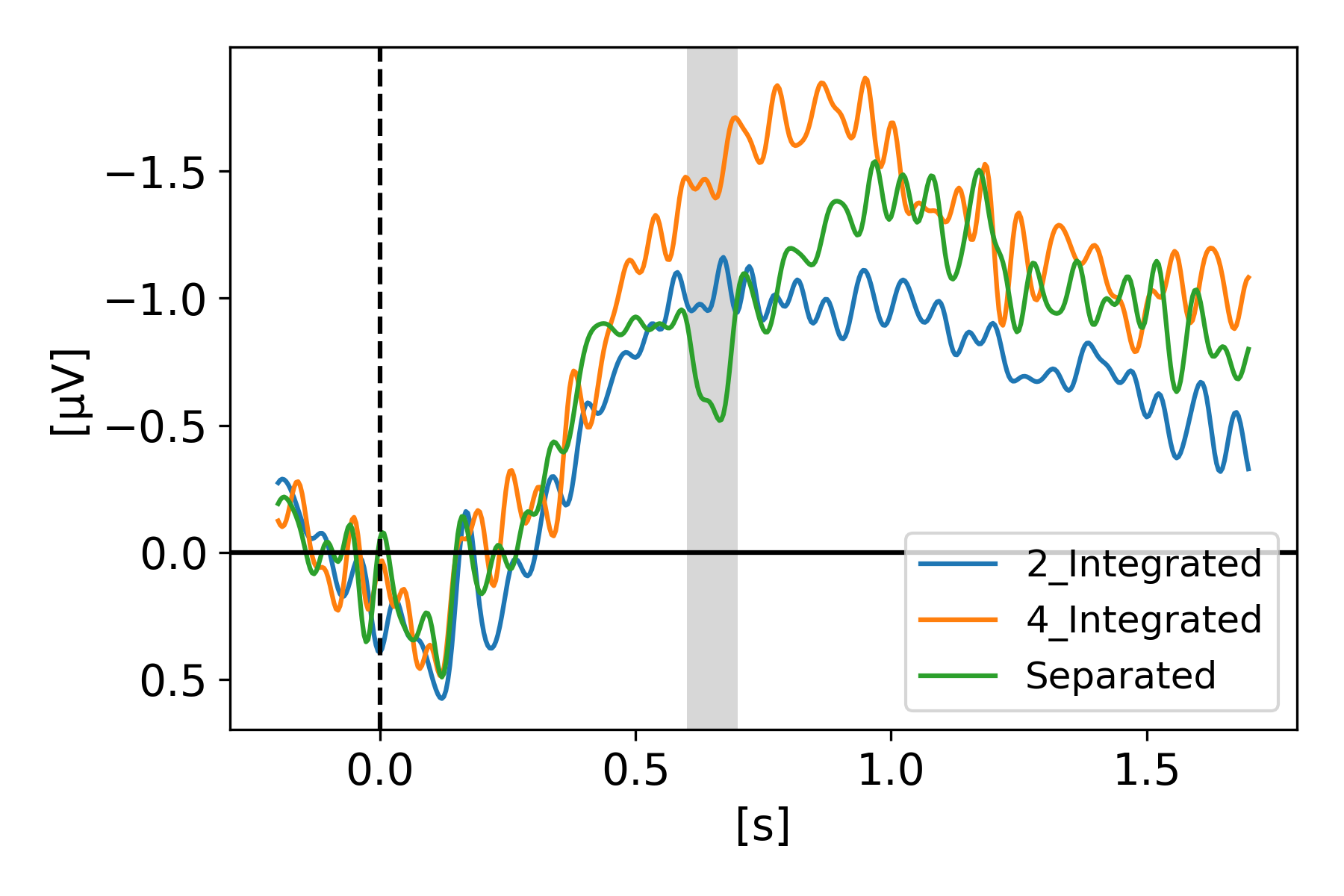}
    \caption{Reproduced results from Balaban et al., 2019 - Experiment 2}
    \label{Balaban-2019-Exp2-reprod}
\end{figure}

\subsection{Gunseli et al., 2019}
\begin{figure} [!htb]
\centering
\begin{subfigure}{.49\textwidth}
    \centering
    \includegraphics[width=.95\linewidth]{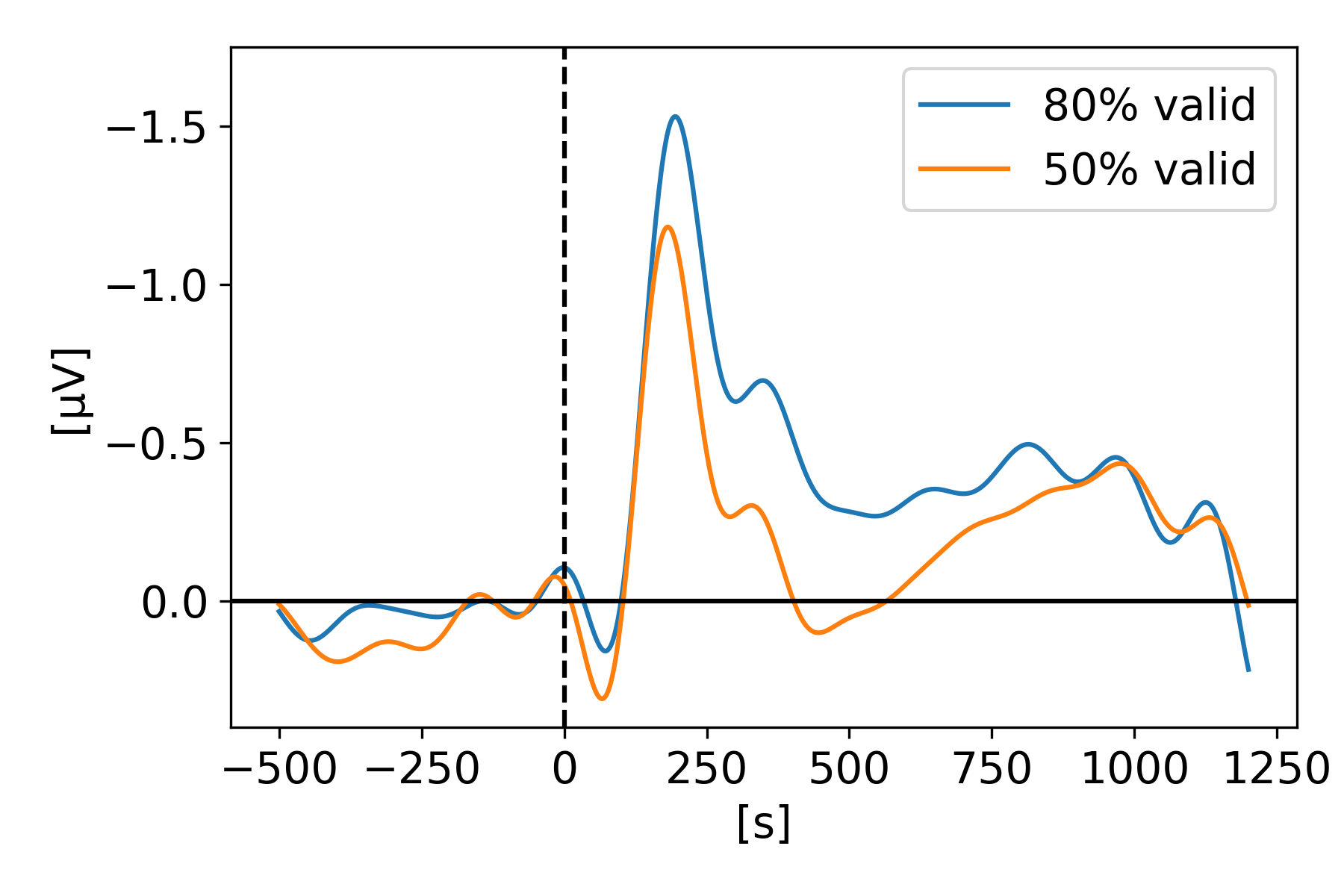}
\end{subfigure}
\begin{subfigure}{.49\textwidth}
    \centering
    \includegraphics[width=.95\linewidth]{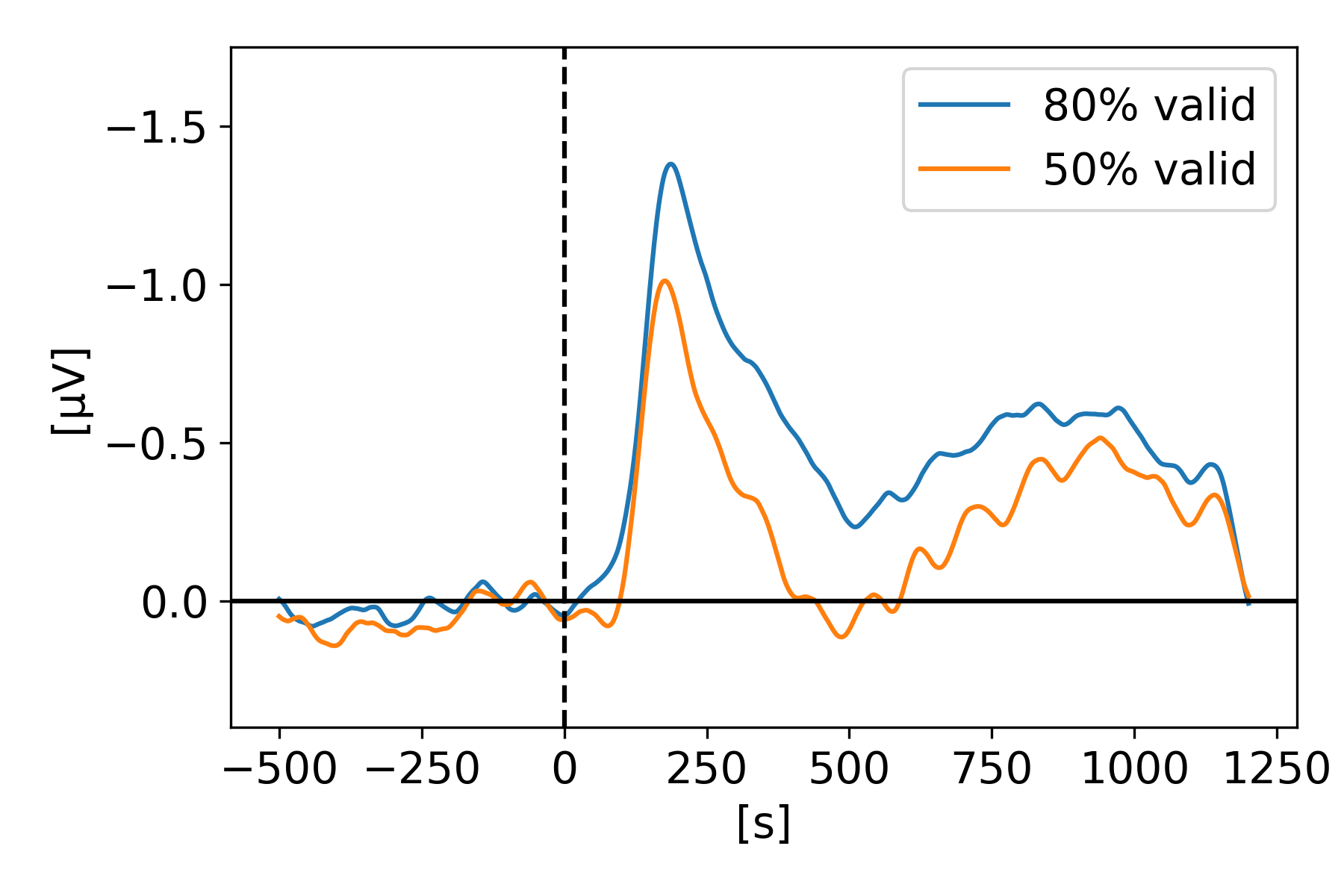}
\end{subfigure}
\caption{Reproduced results from Gunseli et al., 2019. (a) CDA calculated from raw EEG files. (b) CDA obtained from preprocessed MATLAB files.}
\label{Gunseli-2019-reprod}
\end{figure}

In \textit{EEG dynamics reveal a dissociation between storage and selective attention within working memory} (\cite{balaban2019neural}) Gunseli and colleagues tested the hypothesis that within WM, selectively attending to an item and stopping storing other items are independent mechanisms. In order to make participant drop items from WM, they used a retro-cue to indicate which of the items were the target(s). As opposed to identifying the targets at the beginning of the trial like in most WM studies, here they showed 3 items (bars with different orientations) and then, 1s later they showed a retro-cue indicating which item is most likely to be tested, or probed, after the retention phase. Their hypothesis was that if the retro-cue is reliable, the participant would drop the other item(s) creating an imbalance between hemispheres and therefore increasing the CDA signal. Whereas if the retro-cue is not reliable the participants would not drop the item(s) and therefore resulting in a smaller CDA signal. On Figure \ref{Gunseli-2019-reprod}, we see that indeed the CDA is of higher amplitude for trials where the retro-cue is valid 80\% of the time in comparison to the other condition where the retro-cue is valid only 50\% of the time. The right graph was generated using their MATLAB preprocessed data files (.mat) and the left graph shows our reproduced version from their raw EEG data, with a lowpass filter at 6Hz, as they mention in their paper to leave the alpha band out of the CDA signal. Unfortunately, the preprocessing code used to generate the MATLAB preprocessed files was not available and it seems like these files benefited from additional manual cleaning because when we use a higher filter (e.g. 20Hz or 30Hz) with the same pipeline as other studies, both signals end up with a similar amplitude and the effect isn't visible anymore because of a high variability on both signals (50\% vs 80\%). In their paper they mentioned that using a higher filter (e.g. 40Hz) didn't change the results of the statistical analysis (relative to the 6Hz filter). We did not perform any statistical analysis, however, we were only able to obtain a visible CDA difference when plotting the grand average with heavy filtering (6Hz), which helps reduce the variability. Only the PO7-PO8, P7-P8, and O1-O2 electrode pairs were used for generate the CDA.

\subsection{Villena-Gonzalez et al., 2019}
\begin{figure} [!htb]
    \centering
    \includegraphics[width=0.5\textwidth]{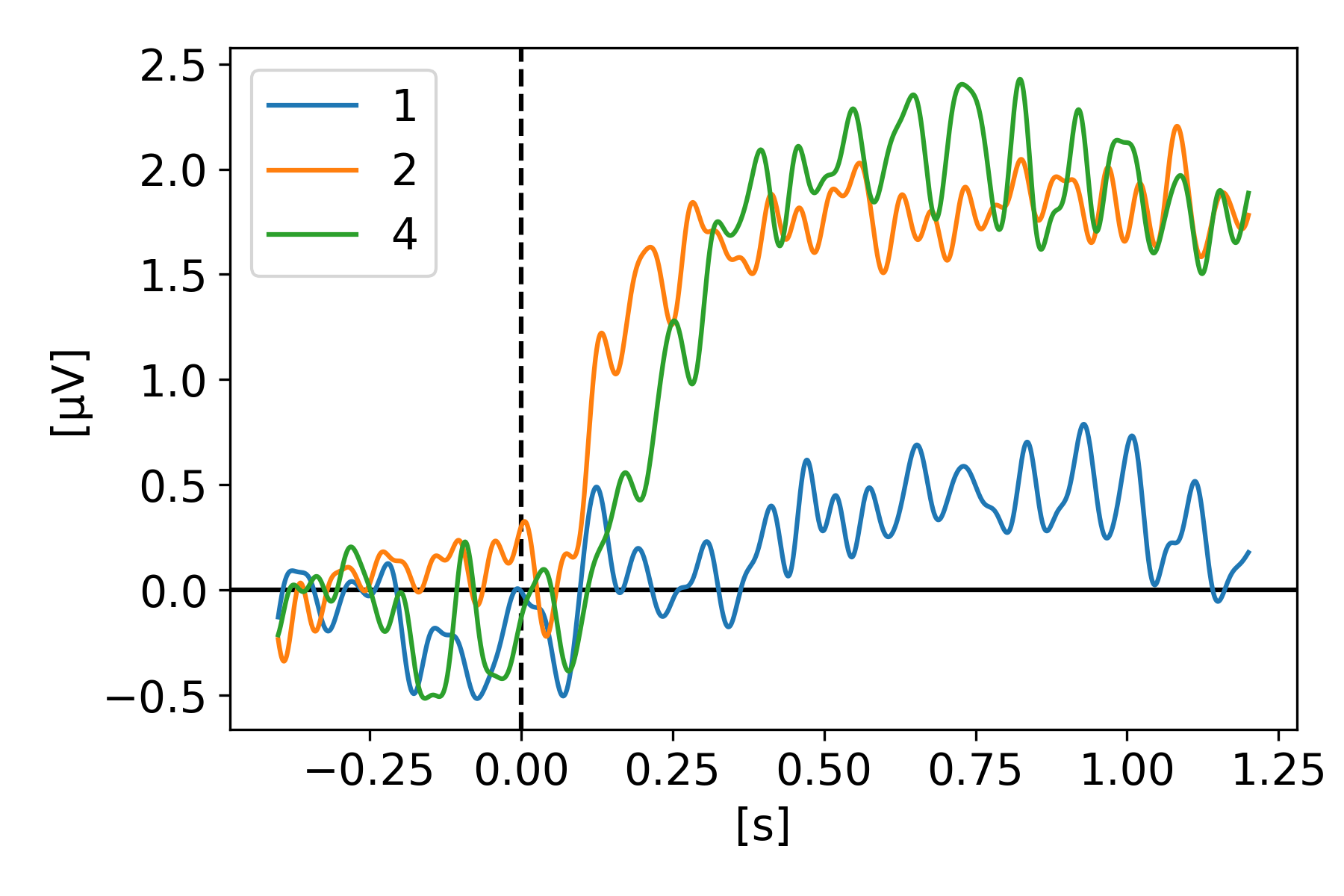}
    \caption{Reproduced results from Villena-Gonzalez et al., 2019}
    \label{Villena-Gonzalez-2019-reprod}
\end{figure}

In \textit{Data from brain activity during visual working memory replicates the correlation between contralateral delay activity and memory capacity} (\cite{villena2020data}), Villena-Gonzalez and colleagues replicated the results from Vogel, 2004 showing that the amplitude of the CDA correlates with the number of items held in WM using a change detection task with set sizes of one, two and four. Moreover, they also looked at the individual performances and showed that participants with higher WM capacity (i.e. better performance on the task) also had a higher CDA amplitude. Figure \ref{Villena-Gonzalez-2019-reprod} shows a clear difference between one target and two or four targets, however, we see similar amplitudes for two and four targets. Unfortunately, we were not able to reproduce their results showing a clear difference of amplitude between two and four targets. In their paper, they had a significant higher amplitude for four targets, which we failed to reproduce after a few attempts at with different automated preprocessing pipelines. Our version doesn't invalidate their conclusion as we also see a higher CDA amplitude the larger the set size, however the effect between two and four isn't as strong as in their findings. This might indicate that the results could have benefited from extra manual cleaning. The following electrode pairs were used to obtain the CDA: TP7-TP8, CP5-CP6, CP3-CP4, CP1-CP2, P1-P2, P3-P4, P5-P6, P7-P8', P9-P10, PO7-PO8, PO3-PO4, and O1-O2.

\subsection{Hakim et al., 2019}
\begin{figure} [!htb]
    \centering
    \includegraphics[width=0.5\textwidth]{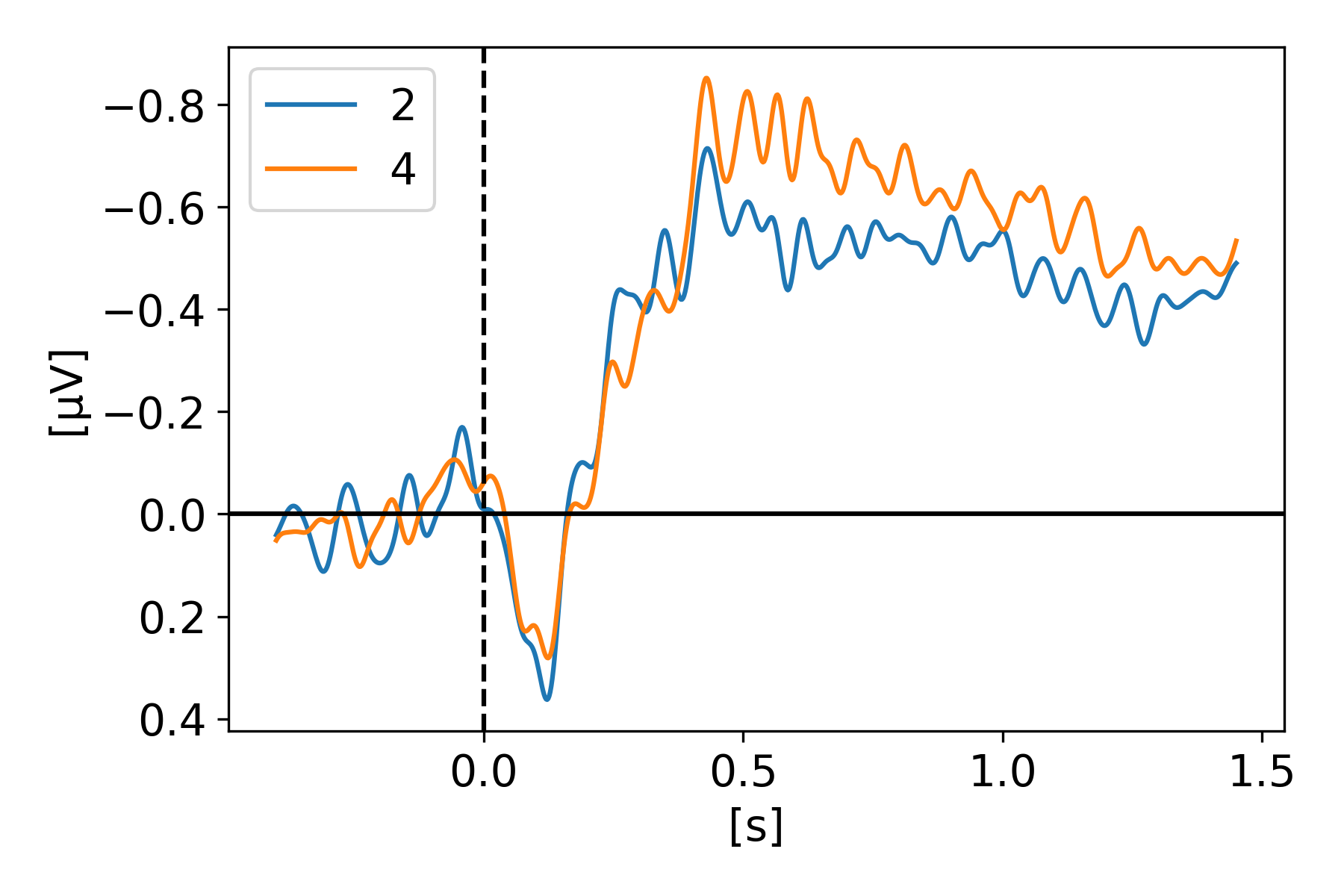}
    \caption{Reproduced results from Hakim et al., 2019}
    \label{Hakim-2019-reprod}
\end{figure}

In \textit{Dissecting the Neural Focus of Attention Reveals Distinct Processes for Spatial Attention and  Object-Based Storage in Visual Working Memory} (\cite{hakim2019dissecting}), Hakim and colleagues showed that the focus of attention in WM is not a monolithic construct but rather involves at least two neurally separable processes: (a) attention to regions in space and (b) representations of objects that occupy the attended regions. On Figure \ref{Hakim-2019-reprod}, the CDA is clearly visible, showing a slightly higher amplitude for a set size of 4 targets vs 2.
The full 1.45s epoch of the change detection task is displayed but excludes the answer part which took place right after. It is important to note that the graph represent the aggregation of 4 sub-experiments with slight variations on the task. The CDA was obtained by using only O1-O2, PO3-PO4, PO7-PO8, P3-P4, and P7-P8 electrode pairs. 

\subsection{Feldmann-Wüstefeld et al., 2018}
\begin{figure} [!htb]
    \centering
    \includegraphics[width=0.75\textwidth]{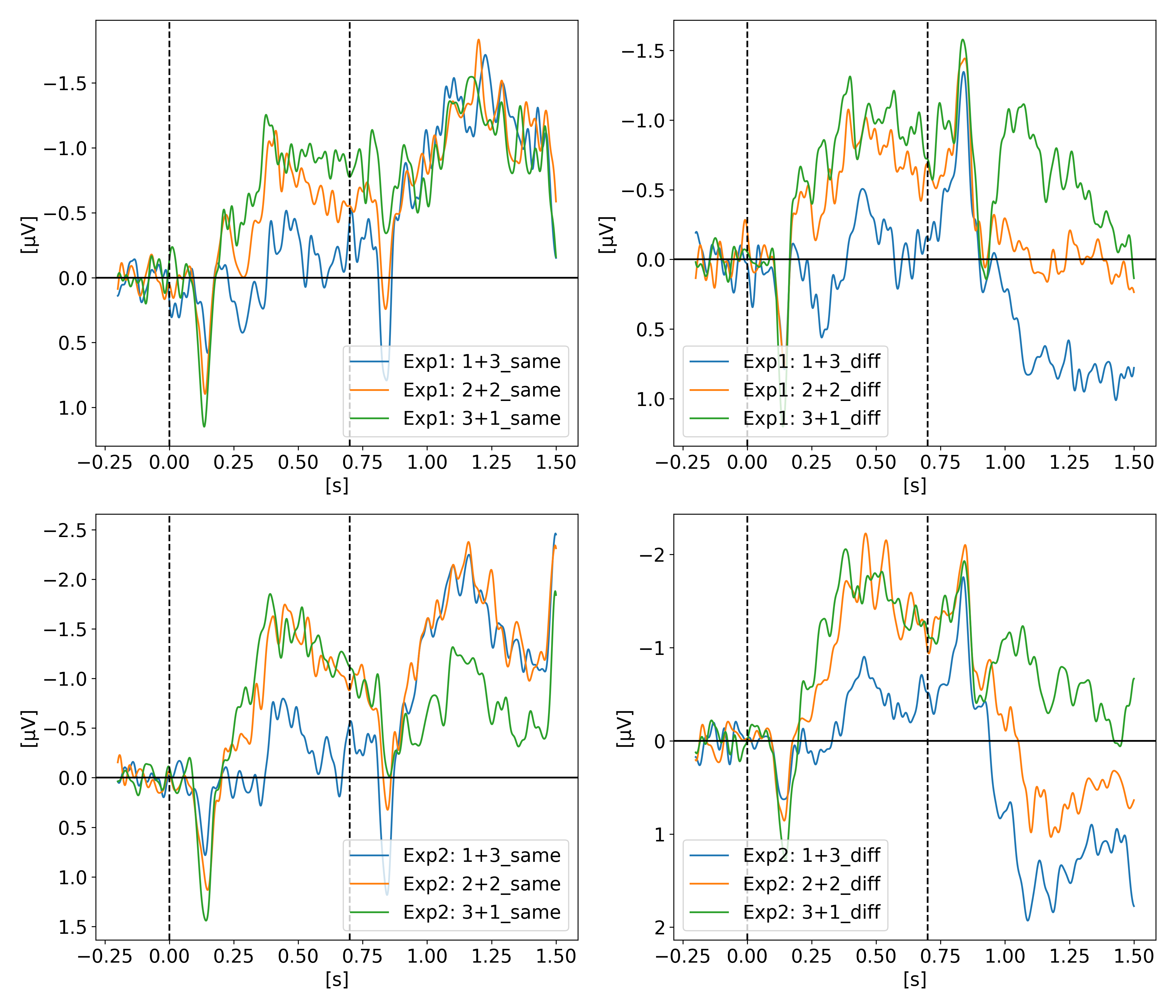}
    \caption{Reproduced results from Feldmann-Wüstefeld et al., 2018}
    \label{Feldmann-Wustefeld-2018-reprod}
\end{figure}

In \textit{Contralateral Delay Activity Indexes Working Memory Storage, not the Current Focus of Spatial Attention} (\cite{feldmann2018contralateral}), Feldmann-Wüstefeld and colleagues seek to explore the recent hypothesis from Berggren and Eimer suggesting that the CDA tracks the current focus of spatial attention as opposed to working memory storage (\cite{berggren2016does}). Figure \ref{Feldmann-Wustefeld-2018-reprod} shows the CDA results of both their experiments in which they displayed four targets among distractors via two sequential memory displays in a change detection task. The first set of targets is shown at t=0s for 200ms then disappears for 500ms after which a second set of targets and distractors appear for 200ms and then disappear for another 500ms after which the probe display is shown for the participant to confirm if the items currently displayed are the same as the targets. The whole 1.4s epoch is displayed on Figure \ref{Feldmann-Wustefeld-2018-reprod}, leaving the answer part out of that figure. A total of four targets were always shown to the participant. The first memory display (i.e. the first set of targets and distractors to be shown) could either have 1, 2 or 3 targets and the second memory display, 700ms later, could add 3, 2 or 1 targets for a combined total of 4. The targets could be either added in the same hemifield or the different (i.e. opposite) hemifield. The expected result is a higher CDA when the targets are added in the same hemifield as this will increase the "net load" (term they will later use in their 2020 paper) and a lower CDA amplitude when added in the opposite hemifield because it would then decrease the net load. Experiment 1, on the upper row, shows such expected results somewhat perfectly as on the top left graph we see all three CDA signals reaching the same amplitude after both sets of targets are added and equals to four items in the same hemifield. As expected, the top right graph shows a slight decrease on the CDA amplitude for \textit{3+1 diff} but more interestingly, a CDA of pretty much zero for the \textit{2+2 diff} condition where two targets were shown in both hemifields cancelling out the CDA signal. For the \textit{1+3 diff} condition, we see the CDA flipping side after the new set of three targets is shown on the opposite side. Experiment 2 was similar to Experiment 1 except for the probe display when the participant provides the answer. In Experiment 1 the probe display showed the items at the same spatial location that they were displayed in either the memory display 1 (at t=0ms) or 2 (at t=700ms). However, in Experiment 2 the probe display was modeled after Berggren and Eimer (\cite{berggren2016does}) experiment such that the items were spatially translated and interleaved. The authors suggested that given that the retention part of Experiment 1 and 2 were identical (i.e. the 1.4s displayed on the graph) and that only the probe display was different, the differences observed between both experiment can only be explained by different memory strategies. They therefore suggested that because the mental representation in Experiment 2 is more difficult as the items are not displayed "as is" but translated on the probe display, the participants most likely transferred items of display 1 (i.e. first set of targets) from working memory (WM) into long-term memory (LTM) therefore explaining why in Experiment 2 the CDA seems more affected by the second set of items rather than equally affected from the first and second set of targets as in Experiment 1. A clear example of that difference between experiments is the \textit{2+2 diff} condition where the CDA is pretty much zero in Experiment 1 but biased towards the second set of targets in Experiment 2. Only the PO7-PO8 electrode pair was used for the CDA signal.

\subsection{Adam et al., 2018}
\begin{figure}
\centering
\begin{subfigure}{.49\textwidth}
    \centering
    \includegraphics[width=.95\linewidth]{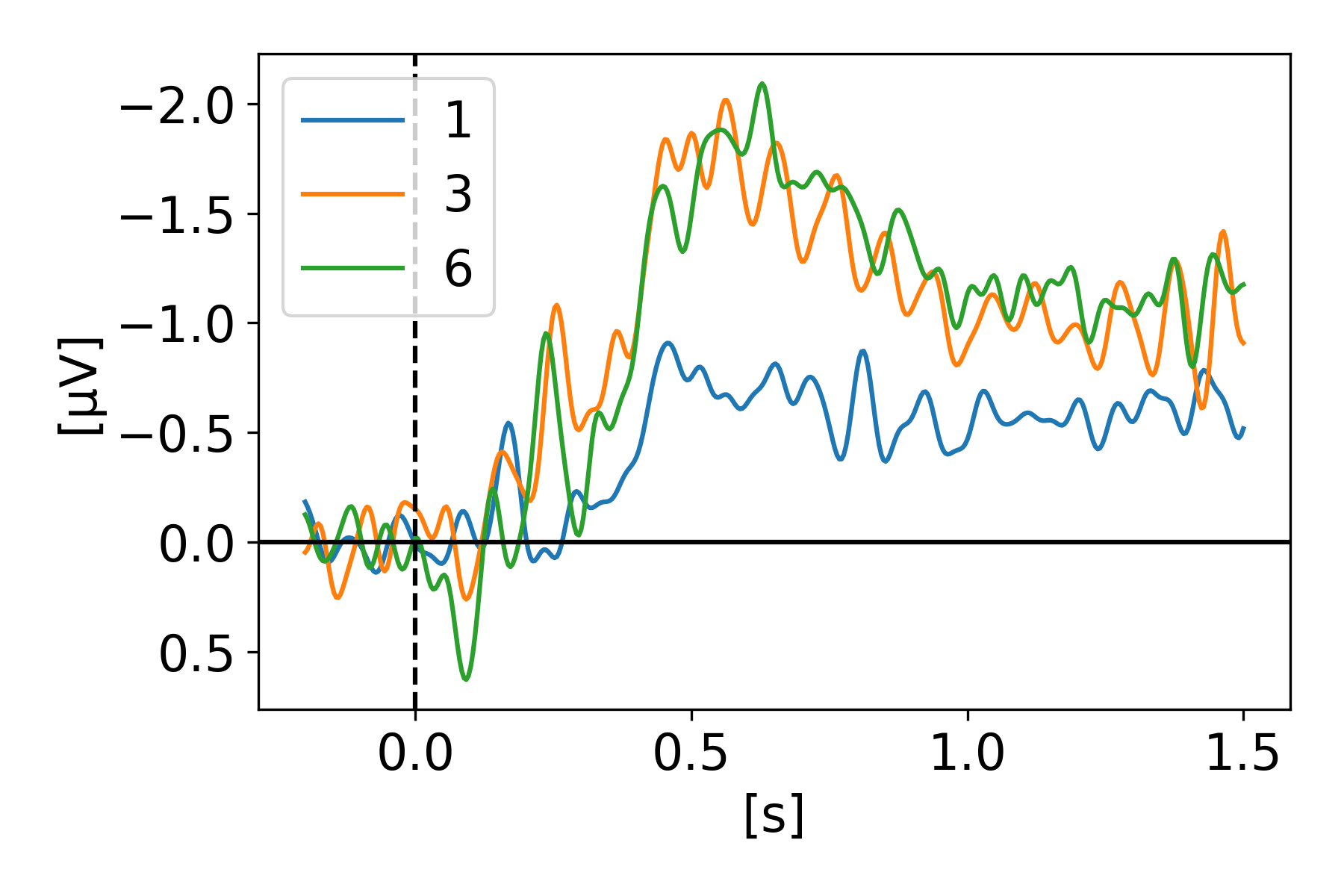}
\end{subfigure}
\begin{subfigure}{.49\textwidth}
    \centering
    \includegraphics[width=.95\linewidth]{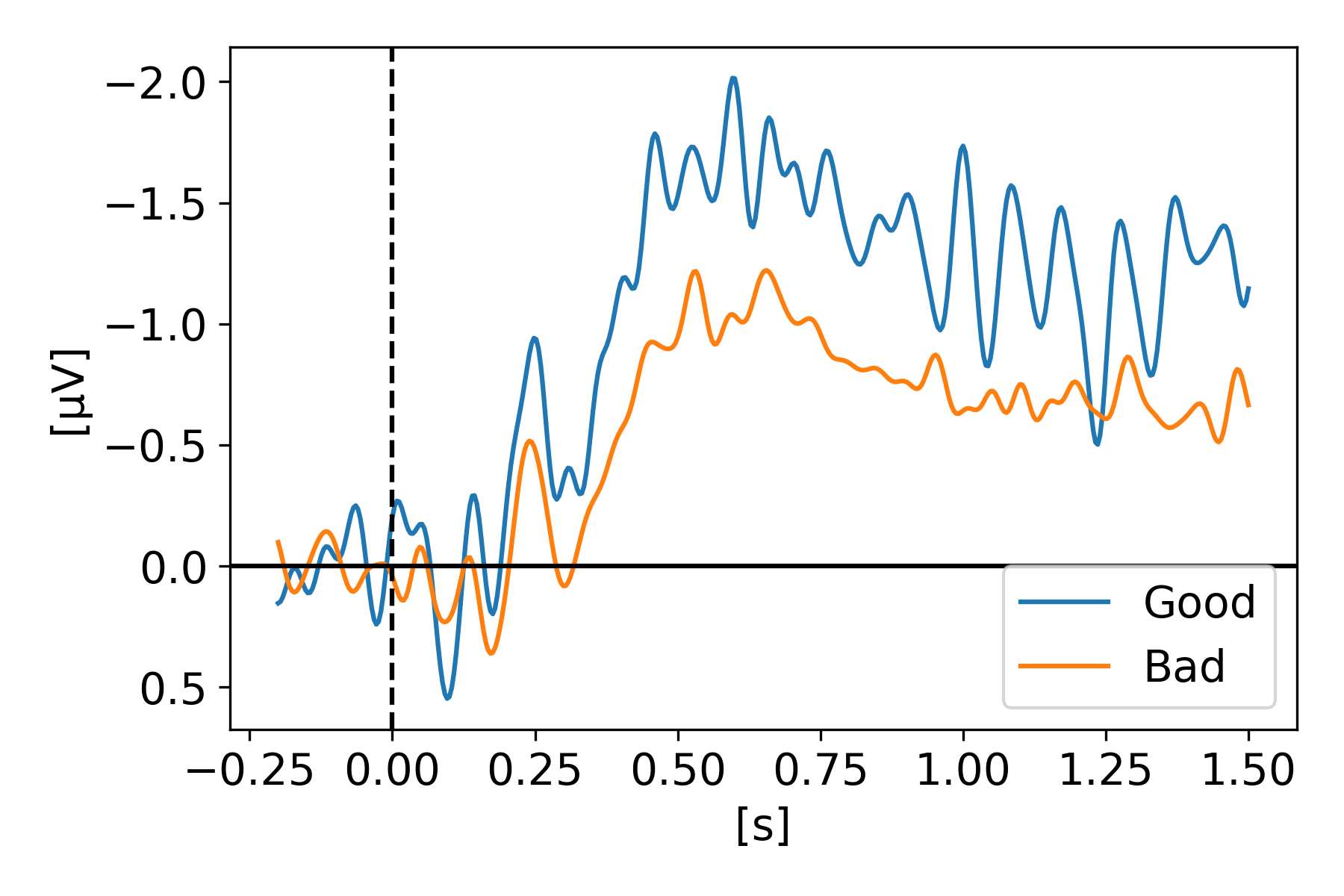}
\end{subfigure}
\caption{Reproduced results from Adam et al., 2018 - Experiment 1 (left) \& 2 (right)}
\label{Adam-2018-reprod}
\end{figure}

In \textit{Contralateral delay activity tracks fluctuations in working memory performance} (\cite{adam2018contralateral}), Adam and colleagues looked at the relationship between the CDA amplitude and working memory performance. Their hypothesis was that if working memory failures are due to decision-based errors and retrieval failures, CDA amplitude would not differentiate good and poor performance trials when load is held constant. If failures arise during storage (i.e. retention phase), then CDA amplitude should track both working memory load and trial-by-trial performance. Their Experiment 1, shown on Figure \ref{Adam-2018-reprod}, showed that the CDA amplitude increased with set size but plateaued at three targets showing a similar amplitude for three or six targets. In their Experiment 2, they kept the set size at 6 and compared the good trials (accuracy of 3 or more targets identified correctly out of 6) vs bad trials (accuracy of 2 targets or less identified correctly). Figure \ref{Adam-2018-reprod} (on the right) shows that indeed the amplitude of the CDA correlated with the performance. The O1-O2, OL-OR, P3-P4, PO3-PO4, T5-T6 electrode pairs were used for the CDA.

\section{Results}

Before analysing the CDA and drawing conclusions on the underlying cognitive functions, we should understand how to best obtain a clean CDA signal in the first place. In this section, we will first discuss what channels and reference(s) various groups have used to obtained their CDA. Second we will discuss some trends we have observed across the reviewed studies.

Note that we did not look at any eye-tracking data and assumed that the subjects respected the instructions of fixating the middle of the screen. Many studies had the eye-tracking data available, however we did not use it nor have we excluded any trials based on eye-movement. This does certainly impact the results we obtained when compared to the original curves of the authors in their paper. Given that such data was not available for all datasets, we decided to not consider it at all.

\subsection{EEG Channels}
The contralateral delay activity (CDA), as its name implies, is a difference in the activity between the left and right hemisphere. The signal is obtained by subtracting one or more contralateral channels to equivalent ipsilateral channel(s) to the attended side. Unsworth et al., mentioned in their 2015 paper that it is now standard procedure for measuring the CDA to use posterior parietal, lateral occipital and posterior temporal electrodes (PO3, PO4, T5, T6, OL, and OR), citing the work of Vogel \& Machizawa, 2004 and Vogel et al., 2005. Looking at the electrodes used in the reproduced studies from 2018 to 2020, there seems to be a slight change towards favouring PO7/PO8 as the best electrode pair. Table ~\ref{cda-channels-pairs-table} shows the selected channels used by the reviewed studies to obtain the CDA. Adam, 2018, used the recommendations from Unsworth, 2015, without the PO7/8 pair, while all the others included at least the PO7/8 pair. Feldmann-Wüstefeld, 2018 \& 2020, and Hakim, 2020, used only that pair for their CDA results and did not average with any other pairs. Interestingly, in her 2019 study, Hakim had use PO7/8 but also P7/8, P3/4, PO3/4, and O1/2, which were not included in the 2020 analysis. Villena-Gonzalez, 2019 used the most electrode pairs of the reviewed studies, including parietal, occipital, temporal and central sites.

\begin{table}[!htb]
\centering
\begin{tabular}{|| l | l ||} 
 \hline\hline
 \textbf{Study} & \textbf{Channels} \\
    \hline
    Feldmann-Westerfel, 2018  & PO7/8 \\
    \hline
    Feldmann-Westerfel, 2020  & PO7/8 \\
    \hline
    Hakim, 2020 & PO7/8  \\
    \hline
    Gunseli, 2019 & PO7/8, P7/8, O1/2 \\
    \hline
    Balaban, 2019 & PO7/8, P7/8, PO3/4 \\
    \hline
    Adam, 2018  & O1/2, OL/R, P3/4, PO3/4, T5/6 \\
    \hline
    Hakim, 2019  & O1/2, PO3/4, PO7/8, P3/4, P7/8 \\
    \hline
    Villena-Gonzalez, 2019  & TP7/8, CP5/6, CP3/4, CP1/2, \\ & P1/2, P3/4, P5/6, P7/8, P9/10, \\ &  PO7/8, PO3/4, O1/2  \\
 \hline
 \hline
\end{tabular}
\caption{Channel pairs used for the CDA.}
\label{cda-channels-pairs-table}
\end{table}

\begin{figure}[!htb]
\centering
\begin{subfigure}{.49\textwidth}
    \centering
    \includegraphics[width=.95\linewidth]{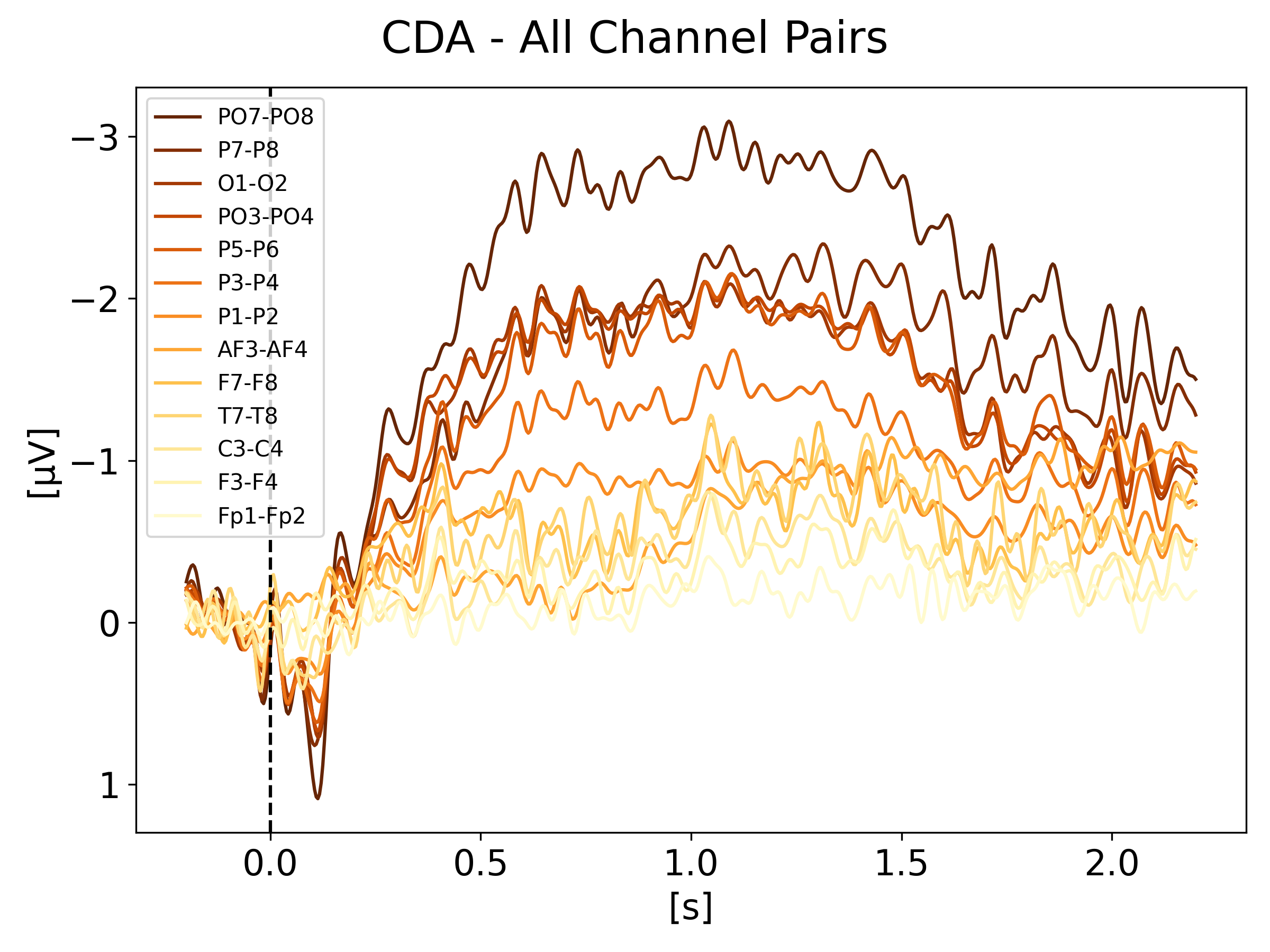}
    \caption{Balaban, 2019}
    \label{channels-balaban2019}
\end{subfigure}
\begin{subfigure}{.49\textwidth}
    \centering
    \includegraphics[width=.95\linewidth]{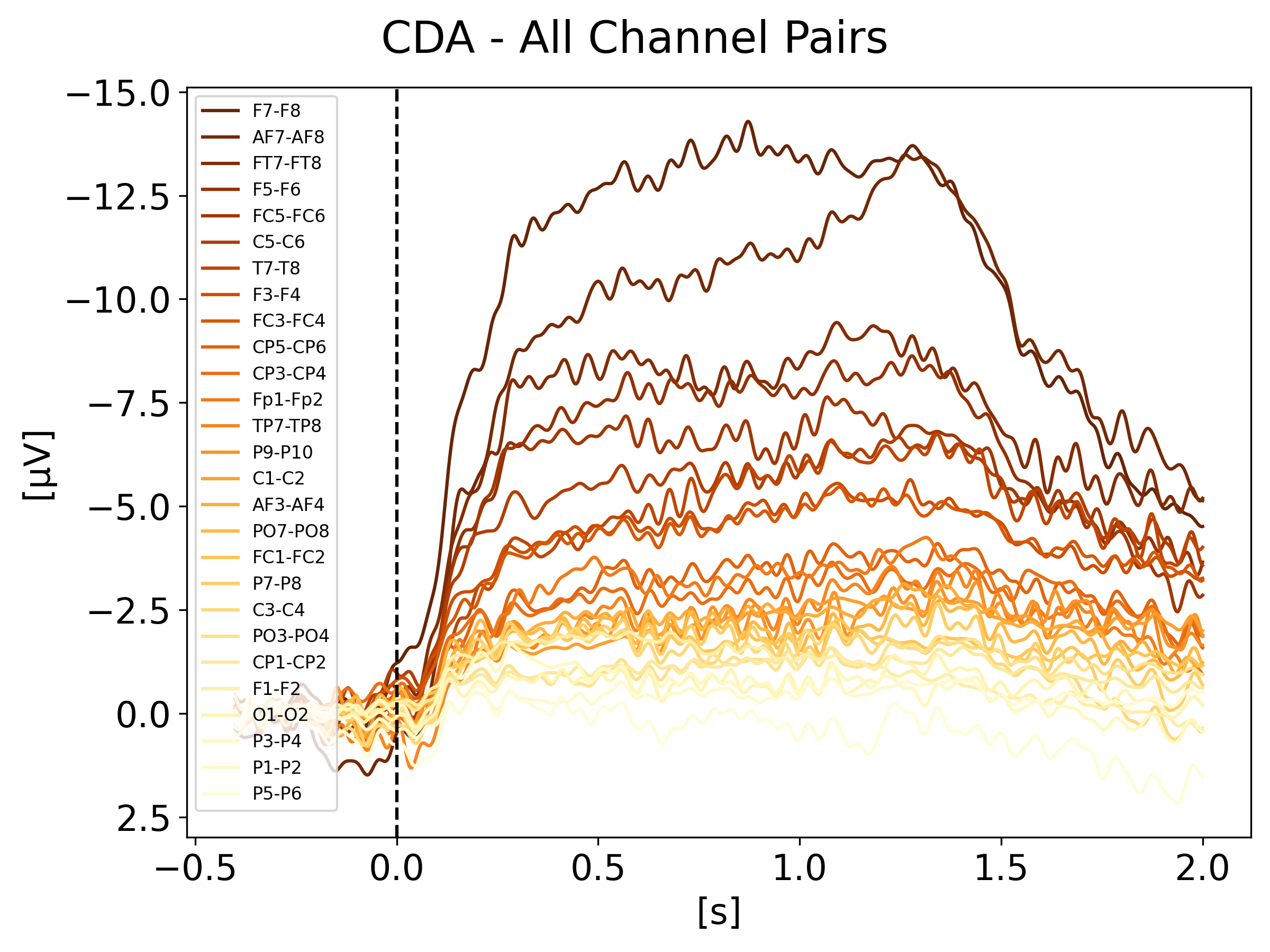}
    \caption{Villena-Gonzalez, 2019}
    \label{channels-villena2019}
\end{subfigure}
\caption{CDA Channel Pairs. (a) Shows the CDA from Balaban et al., 2019 for all available channel pairs. The CDA is the grand average across subjects from all trials with a good performance for the condition integrated shape in experiment 1. (b) Shows the CDA from Villena-Gonzalez et al., 2019 for all available channel pairs. The CDA is the grand average across subjects from all trials with a good performance for the condition with 2 items.}
\label{fig-channels}
\end{figure}

In order to better understand the signal shape and amplitude coming from the various electrode pairs, we looked at each pair for each condition of each study. Regardless of what the study actually used for their analysis, we used all channel pairs available in the raw data. We included here only the breakdown for Balaban 2019 and Villena 2019 on Figures \ref{channels-balaban2019} and \ref{channels-villena2019} respectively. All the other studies followed the trend of Balaban 2019 showing a stronger CDA from the parietal sites with PO7/8 being the best candidate (i.e. strongest CDA amplitude) across studies. Villena 2019, is the only one showing different results with a very strong frontal lateralized activity (see Figure \ref{channels-villena2019}). We initially thought these results were odd, until we analyzed our own data recorded for another project while writing this review, for which we also observed frontal activity being way stronger than parietal activity. Our task is a 3D-MOT task (see \cite{faubert2013professional}). 

Finally, the approach for averaging the channel pairs also varied. Some groups subtracted the pairs first and then averaged across pairs, while others averaged one side first and then subtracted to the average on other side. The order in which the averaging and subtracting happens doesn't really matter if no other operation is done on the signal and if they are all referenced to the same reference. For example, if we are trying to use the pairs PO7-PO8, P7-P8 and PO3-PO4, we could either start by subtracting each pair individually (e.g. signal from PO8 minus signal from PO7) and then averaging the 3 signals obtained, giving us the resulting CDA or we could average one side together (PO7, P7, and PO3) and subtract it from the average signal from the other side (PO8, P8, and PO4). Both approaches would give the same CDA for a given trial.

\subsection{EEG Reference(s)}
The EEG signal being an electric potential difference between two electrodes, namely the electrode of interest and the reference, it is no surprise that the reference plays an important role in EEG studies. Changing the reference can drastically change the signal obtained and therefore influence the conclusion of a study. The ideal reference would be a neutral point with no electrical activity to which we could measure a difference of potential being only the activity of interest. Unfortunately, no such point exists on the body, leaving the problem of finding a good reference unsolved or open to interpretation. Best practices in EEG studies include (1) using the average of left and right mastoids or (2) to re-reference the signal offline to the average of all channels or (3) using one or multiple electrode(s) on the midline such as Cz or (4) using a mathematical reference such as the reference electrode standardization technique (REST) or a Laplacian approach (\cite{yao2019reference}). Here in the reviewed studies, all of them used to the average of left and right mastoids except for Feldmann-Wusterfeld 2018 and 2020 which used the average of all electrodes.

Given that CDA is a difference between left and right hemispheres, the reference isn't as important as in most evoked-related potential (ERP) studies. The choice of reference will impact the visual inspection and the cleaning of the data, however, when it comes to obtaining the CDA itself, because we are subtracting one electrode to the other, the reference gets somewhat cancelled and therefore does not impact the resulting CDA signal as much. For example, if the contralateral electrode is PO8(-Ref) and the ipsilateral electrode is PO7(-Ref), then CDA (contralateral - ipsilateral) is (PO8 - Ref) - (PO7 - Ref) which is equivalent to PO8 - PO7 as the Ref cancels out. It is therefore still important to consider the reference for cleaning the signal but to keep in mind that the CDA itself isn't as affected as much by the reference as typical, non-latelarized, ERPs. This actually makes the CDA signal even more robust to noise, assuming that most EEG noise would be visible on both channels, and be cancelled out during the subtraction. A noise that would survive that subtraction would be a noise seen only by one electrode or one 'side' of the head.



\subsection{CDA Decay}
If the amplitude of the CDA correlates with the number of items being tracked, one could expect the amplitude to remain somewhat stable for the whole retention or tracking duration. However, in all the replicated studies we can observe the amplitude of the CDA declining way before the end of the retention phase even when the participants did not lose track of the item(s).

In later section \ref{cda-performance} we look at the CDA when subjects lost track of one of more item(s) (i.e. bad trials) but all figures previously showed with the reproduced results and discussed in the methods section were generated from \textit{good performances} only, meaning that the subjects did remember all the items for the whole duration of the trial and provided good answers at the end. As we can see on Figures \ref{Feldmann-Wustefeld-2020-reprod}, \ref{Balaban-2019-Exp1-reprod}, \ref{Balaban-2019-Exp2-reprod}, \ref{Gunseli-2019-reprod}, and \ref{Hakim-2019-reprod} the CDA is starting to decrease way before the end of the trials. In Hakim, 2020, they added interruptions trying to interfere with WM processes and CDA indeed dropped significantly shortly after such interruptions. And yet, despite the important drop in amplitude the participants still provided correct answers. While none of the eight studies addressed directly what seems to be a natural CDA decay, some suggested (e.g. Hakim, 2020; Feldmann-Wüstefeld, 2018) that the items could be transferred from short term memory to long term memory therefore affecting the CDA amplitude observed here.

\subsection{Recall} \label{recall}
One interesting phenomena that most studies did not mention in their paper is the recall phase, i.e. when the participant is providing the answer. Interestingly, in most studies, we are observing a re-increase in the CDA amplitude as if the items were brought back to working memory and sometimes the CDA signal is even of higher amplitude than during the identification and retention (or tracking) phases.

\begin{figure} [!htb]
\centering
\includegraphics[width=.5\textwidth]{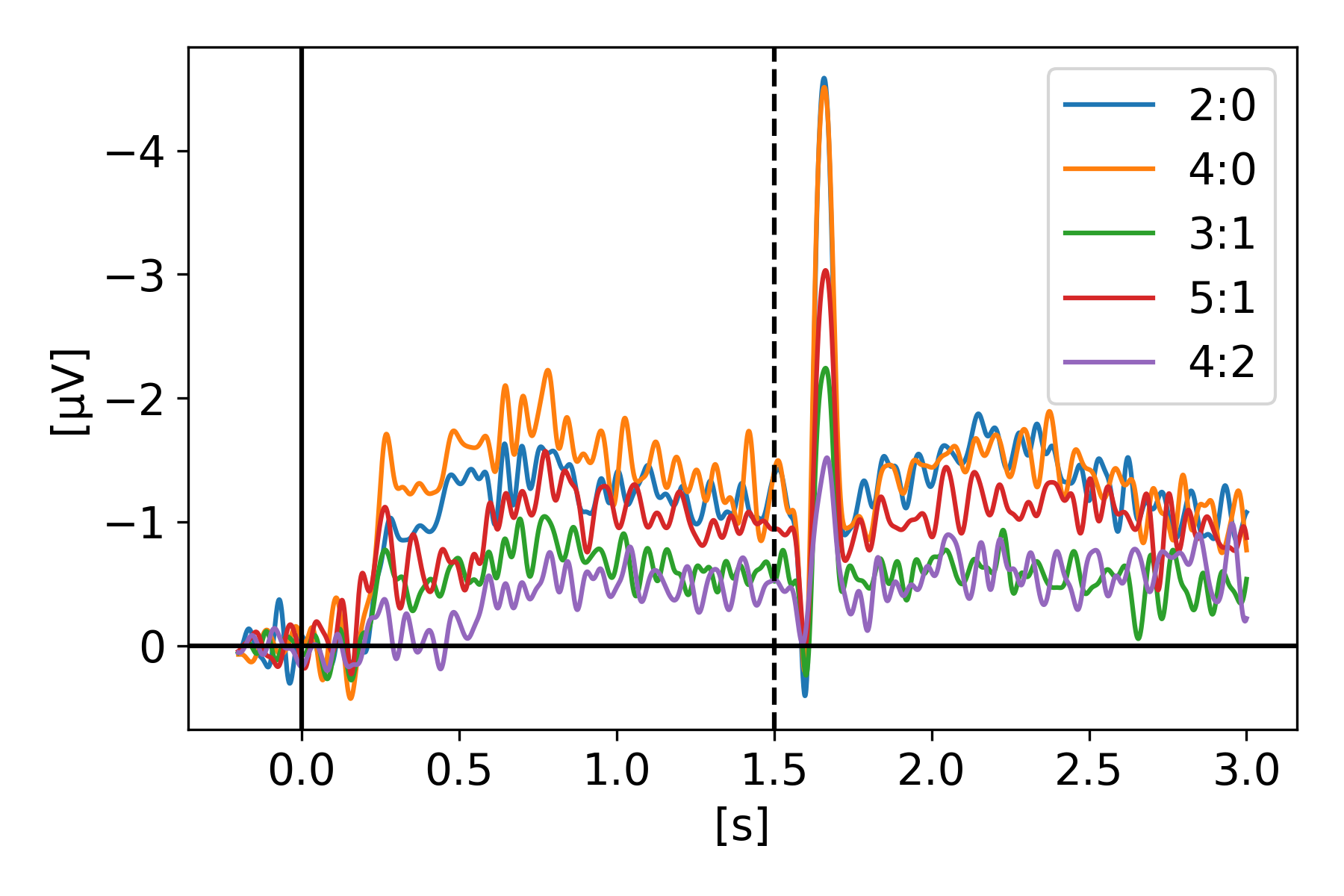}
\caption{Feldmann-Wüsterfeld, 2020 - Recall/Probe. Same CDA as on Figure \ref{Feldmann-Wustefeld-2020-reprod} but with a longer epoch, showing a CDA re-increase during recall (t $>$ 1.5s).}
\label{recall-Feldmann2020}
\end{figure}

\begin{figure} [!htb]
\centering
\includegraphics[width=.8\textwidth]{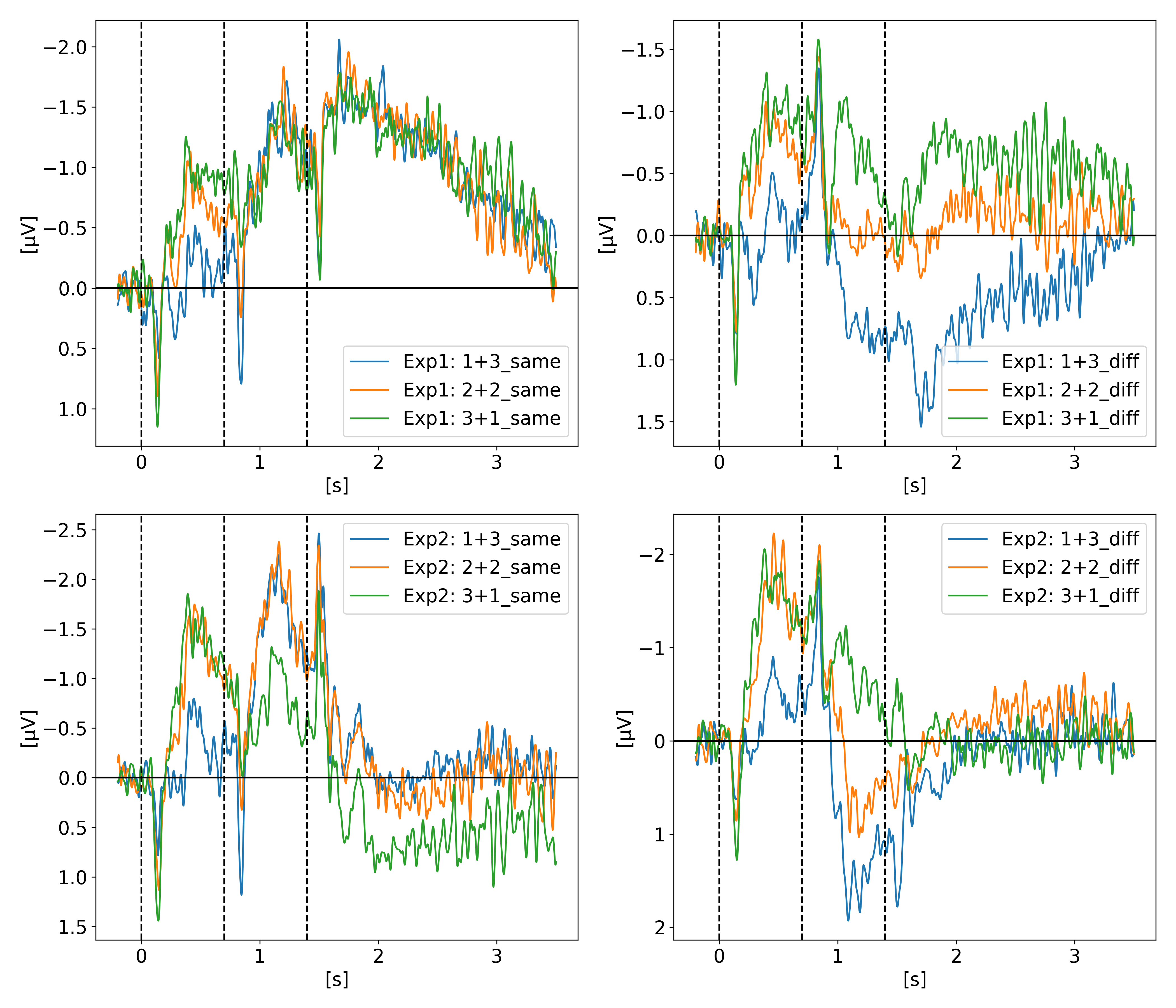}
\caption{Feldmann-Wüsterfeld, 2018 - Recall/Probe. Same CDA as on Figure \ref{Feldmann-Wustefeld-2018-reprod} but with a longer epoch, showing a CDA re-increase during recall (t $>$ 1.5s) for Experiment 1.}
\label{recall-Feldmann2018}
\end{figure}

\begin{figure} [!htb]
\centering
\includegraphics[width=.8\textwidth]{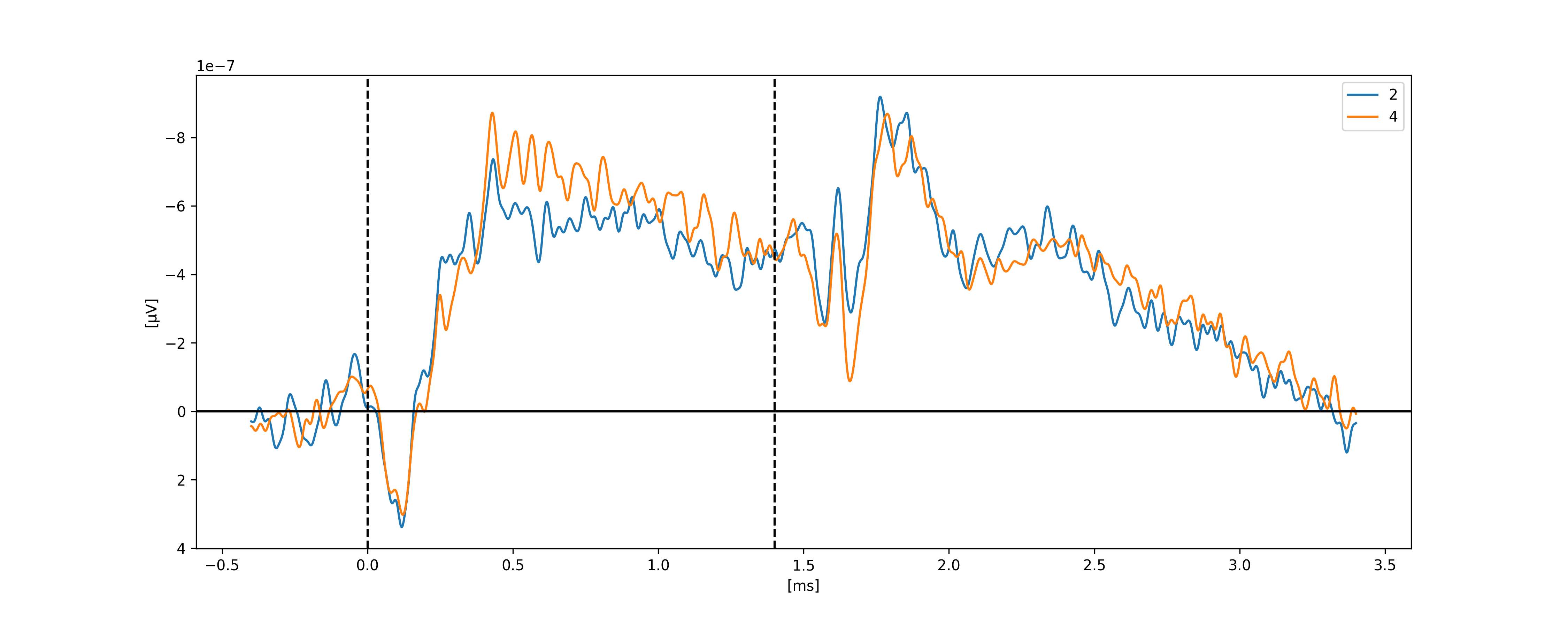}
\caption{Hakim, 2019 - Recall/Probe. Same CDA as on Figure \ref{Hakim-2019-reprod} but with a longer epoch, showing a CDA re-increase during recall (t $>$ 1.4s).}
\label{recall-Hakim2019}
\end{figure}

\begin{figure}
\centering
\begin{subfigure}{.49\textwidth}
    \centering
    \includegraphics[width=.95\linewidth]{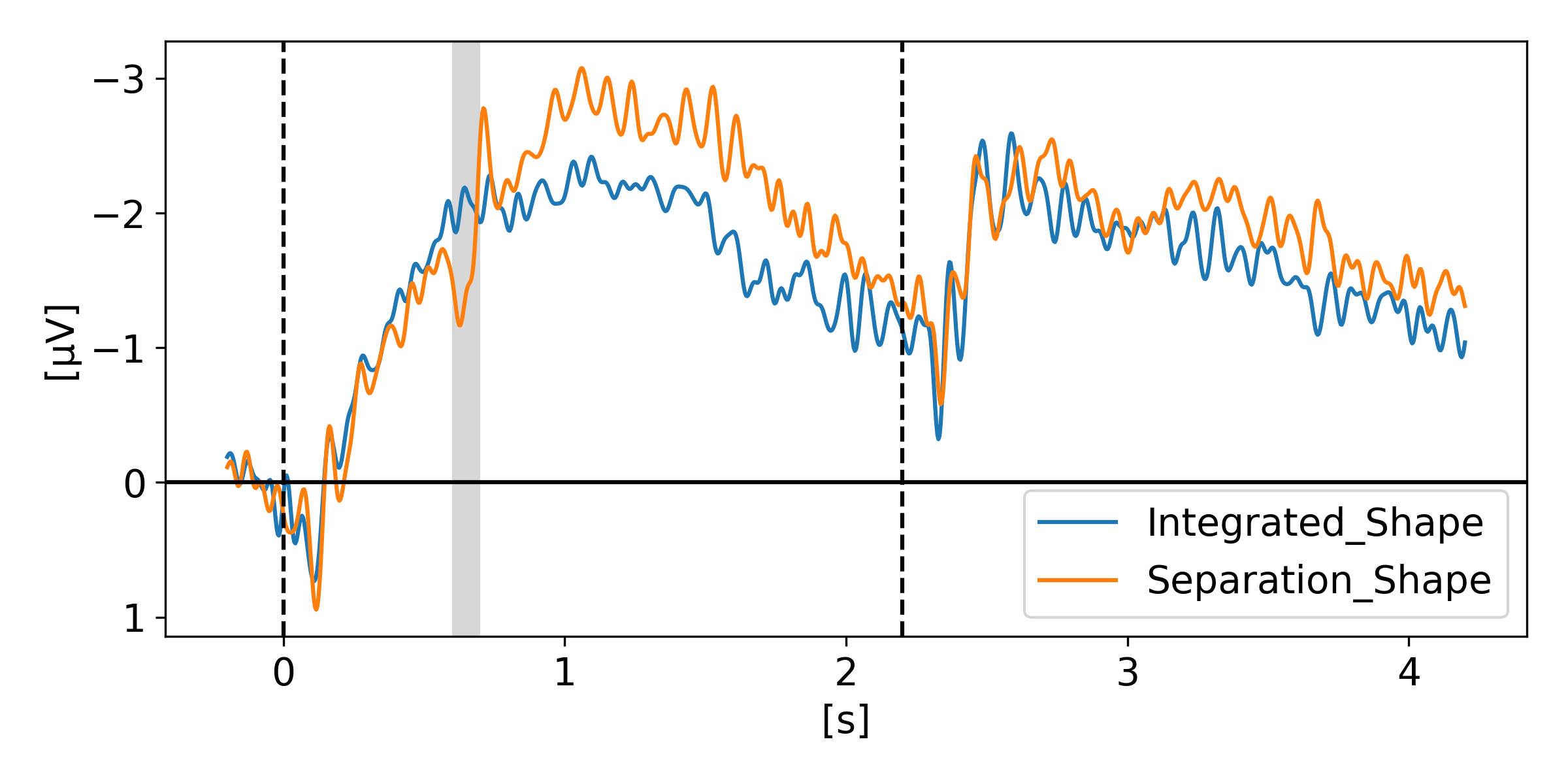}
\end{subfigure}
\begin{subfigure}{.49\textwidth}
    \centering
    \includegraphics[width=.95\linewidth]{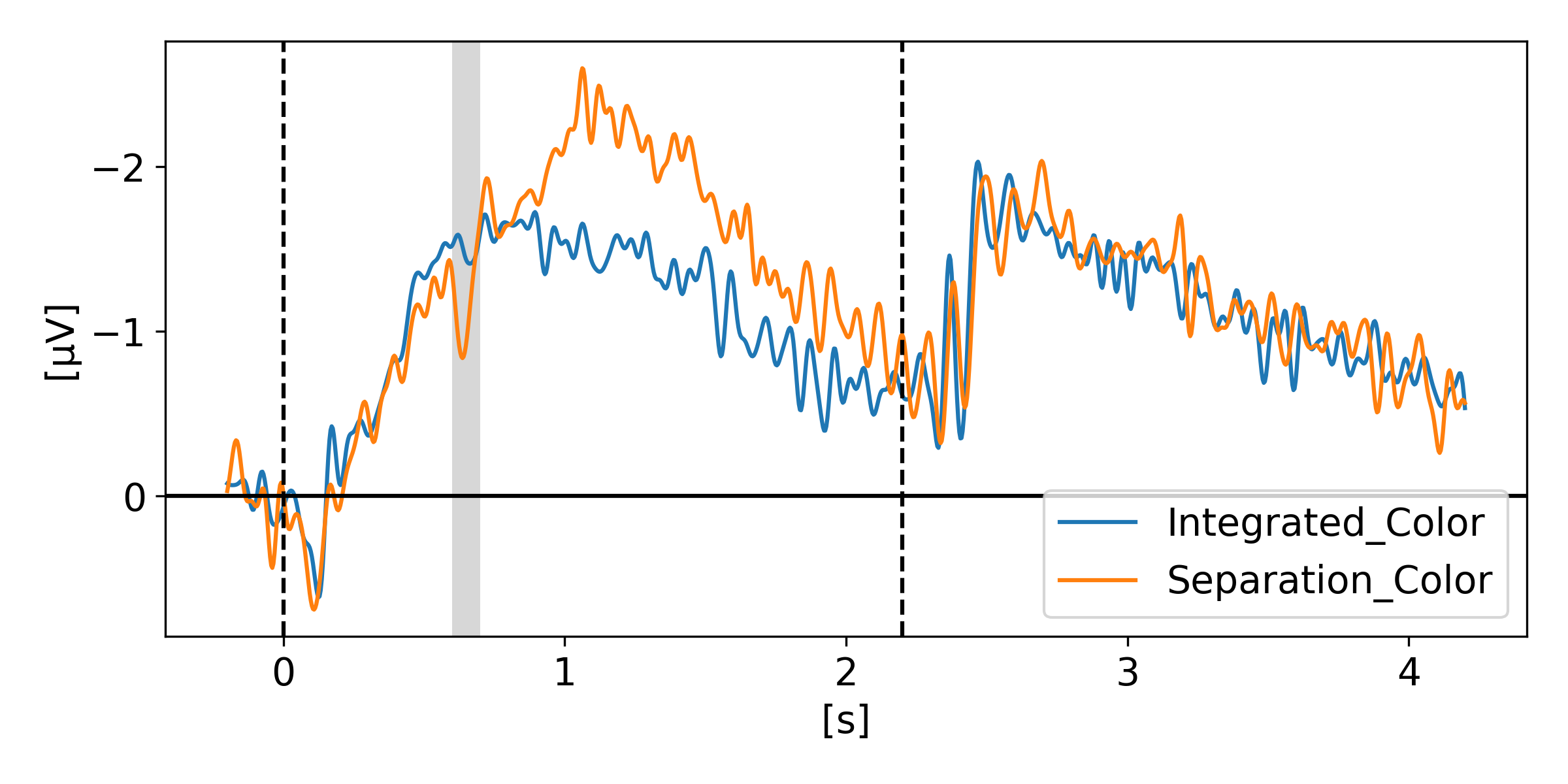}
\end{subfigure}
\caption{Balaban, 2019 - Recall/Probe. Same CDA as on Figure \ref{Balaban-2019-Exp1-reprod} but with a longer epoch, showing a CDA re-increase during recall (t $>$ 2.2s)}
\label{recall-Balaban2019}
\end{figure}

Figures \ref{recall-Feldmann2020}, \ref{recall-Feldmann2018}, \ref{recall-Hakim2019}, and \ref{recall-Balaban2019} show the CDA graphs but with a longer epoch this time, including two seconds after the end of the retention phase. The right-most dashed line represents when the participant was asked to provide an answer. The figures all show a re-increase in the CDA amplitude, also followed by what seems to be the natural CDA decay discussed in the last section.

While we did not investigate this phenomena further, we thought it would be relevant to highlight it in this review as this phase was not looked at in any of the reproduced studies and could help shed light on the neural correlates involved in VMW and underlying the CDA. For example, in their 2018 study, Feldmann-Wüstefeld and colleagues mentioned that the difference between their Experiment 1 and Experiment 2 is the memory strategy the participants might have used. Interestingly, when we compare the recall/answer part of Experiment 1 we see the CDA amplitude re-increasing, however in Experiment 2 we do not see the CDA re-increasing, as if indeed a different recall strategy was used. Only two studies did not show such re-increase of CDA amplitude during recall: Villena-Gonzalez, 2019 and Gunseli, 2019. The graphs are available online in our repository with the analysis.


\subsection{CDA Amplitude vs Number of Items} 
\label{cda-items}
Looking at the reproduced results, it seems fair to conclude that indeed the amplitude of the CDA correlates with the number of items tracked by the participants up to a plateau of 3 to 4 items. This correlation and plateau has been discussed in previous CDA studies (e.g. \cite{vogel2004neural, unsworth2015working, luria2016contralateral}). Figure \ref{Villena-Gonzalez-2019-reprod} from Villena-Gonzalez, 2019, data shows a clear difference in CDA amplitude between one, two and four items. Figure \ref{Adam-2018-reprod} from Adam, 2018, data shows a clear difference between one and three or one and six but a very similar amplitude for three and six items, aligned with some sort of CDA amplitude plateau around three or four items. Figure \ref{Hakim-2019-reprod} from Hakim, 2019, data shows a small difference between two and four items. 

The work from Feldmann-Wüstefeld, 2018 (Figure \ref{Feldmann-Wustefeld-2018-reprod}) shows that even after the initial identification and tracking phases, the CDA can be increased by adding items on the same side or decreased by adding items on the opposite side. Similarly, the work from Balaban, 2019 (Figures \ref{Balaban-2019-Exp1-reprod} and \ref{Balaban-2019-Exp2-reprod}) shows that when a whole item splits in two separate parts that need to be tracked, the CDA increases. Feldmann-Wüstefeld, 2020 (Figure \ref{Feldmann-Wustefeld-2020-reprod}) used a bilateral task with different loads on both sides creating a net load of either two or four items and what their work shows is that different combinations of net load of two or four produce different CDA amplitudes. The more objects on the opposite side, the lower the CDA amplitude despite the same net load. However, the results are aligned with the CDA theory saying that the more objects the higher the amplitude. The 4:0 condition shows a higher amplitude than the 2:0 condition. The 5:1 shows a higher amplitude than the 3:1.

\subsection{CDA Amplitude vs Individual Performance} 
\label{cda-performance}
If the CDA amplitude correlates with the number of items held in memory, the CDA should be indicative of the performance of a specific trial. This obviously holds true only if the working memory failures occur during the retention phase and not during the recall phase to provide the answer. If such memory fault occurs during the initial identification phase, the CDA would not drop per se but instead reach a lower amplitude than the expected peak amplitude should the participant have tracked the correct amount of items. However, if the participant mistakenly identifies the wrong target (for example in a multiple-object tracking task) and confidently holds and tracks the correct number of items, despite some of them being wrong, this kind of mistake would result in the same CDA amplitude as if the right items were kept in memory.
While the current state of CDA literature lacks the trial-by-trial analysis to evaluate the role CDA plays in performance, several studies have looked at the individual differences in CDA amplitude vs working memory performance or capacity. For example, Unsworth and colleagues in 2015 (\cite{unsworth2015working}) replicated the work from Vogel \& Machizawa, 2004 (\cite{vogel2004neural}) showing a correlation between CDA and performance on a change detection task where high working memory individuals had larger CDA (\textit{r}=0.30; Unsworth, 2015) and concluded their study saying that CDA is a reliable and valid individual measure of working memory that predicts behavioral performance on visual working memory tasks. Of the studies reviewed here, only three looked at the direct correlation between individual working memory performance and CDA amplitude. Most studies looked at the relationship between task conditions and performance, as well as the relationship between task conditions and CDA amplitude, however, only three looked at the direct correlation between performance and CDA amplitude (Adam, 2018; Feldmann-Wüstefeld, 2020; Villena-Gonzalez, 2019). Among these three studies, working memory performance, or capacity, was evaluated differently based on the task and the granularity in the answers provided by the participants.

In Adam, 2018, they used a whole report task with a set size of six items and noted the participants' answer accuracy between zero and six targets identified correctly. With this granularity, they looked at the correlation between CDA amplitude and accuracy (from 0/6 to 6/6, steps of 1) and showed individual differences in working memory performance with a correlation of \textit{r}=-.26 and \textit{p}=.028. They also noted that the magnitude of the effect is relatively small, but consistent with previously observed effects in the literature, citing Unsworth, 2015 work. With their EEG and behavioural data we were able to reproduce a similar result with a slightly higher and more significant correlation (\textit{r}=-0.34, p=0.001). It is important to note that the classes are quite unbalanced as there are very few trials with an accuracy score of 0/6 (all wrong) or 6/6 (all good) and two thirds (66\%) of the trials have an accuracy of either 2/6 or 3/6.

In Feldmann-Wüstefeld, 2020, they used the K score from \cite{cowan2001magical} as a performance measure for each subject, where \textit{K = N × (hit rate — false alarm rate)} and reported a correlation \textit{r}=-0.43, \textit{p}=0.026 in their study. With their EEG and behavioural data we were able to reproduce a similar result with a slightly higher correlation (\textit{r}=-0.53, p=0.016).

Villena-Gonalez 2019 used the formula from \cite{pashler1988familiarity}, \textit{K = S x ((H-F)/(1-F))}, where \textit{H} is the observed hit rate, \textit{F} the false alarm rate and \textit{S} is the higher set size (maximum number of to-be-remembered items). In their study, instead of looking at the correlation with WMC and CDA amplitude, they looked at the increase in CDA amplitude between set size of two and set size of four items and reported \textit{r}=0.448; \textit{p}=0.0159 showing that participants with high WMC (i.e. better performance) showed larger amplitude increase in CDA between two and four items, compared with participants with low WMC. Unfortunately, their behaviour files were not available with their EEG data, preventing us from calculating the WMC with the same formula. In the EEG files, a binary trigger identifying good and bad answers was available so we were able to calculate the accuracy for each participant. Unfortunately, when correlating the difference in amplitude between ss=2 and ss=4 with accuracy for each participant we found no significant correlation and therefore were unable to reproduce the effect of CDA on individual performance.

The figures of the correlations, statistical analyses and distributions are available in the supplementary material.


\subsection{Subjects Variability} 
\label{subjects-variability}
\begin{figure} [!ht]
    \centering
    \includegraphics[width=\linewidth]{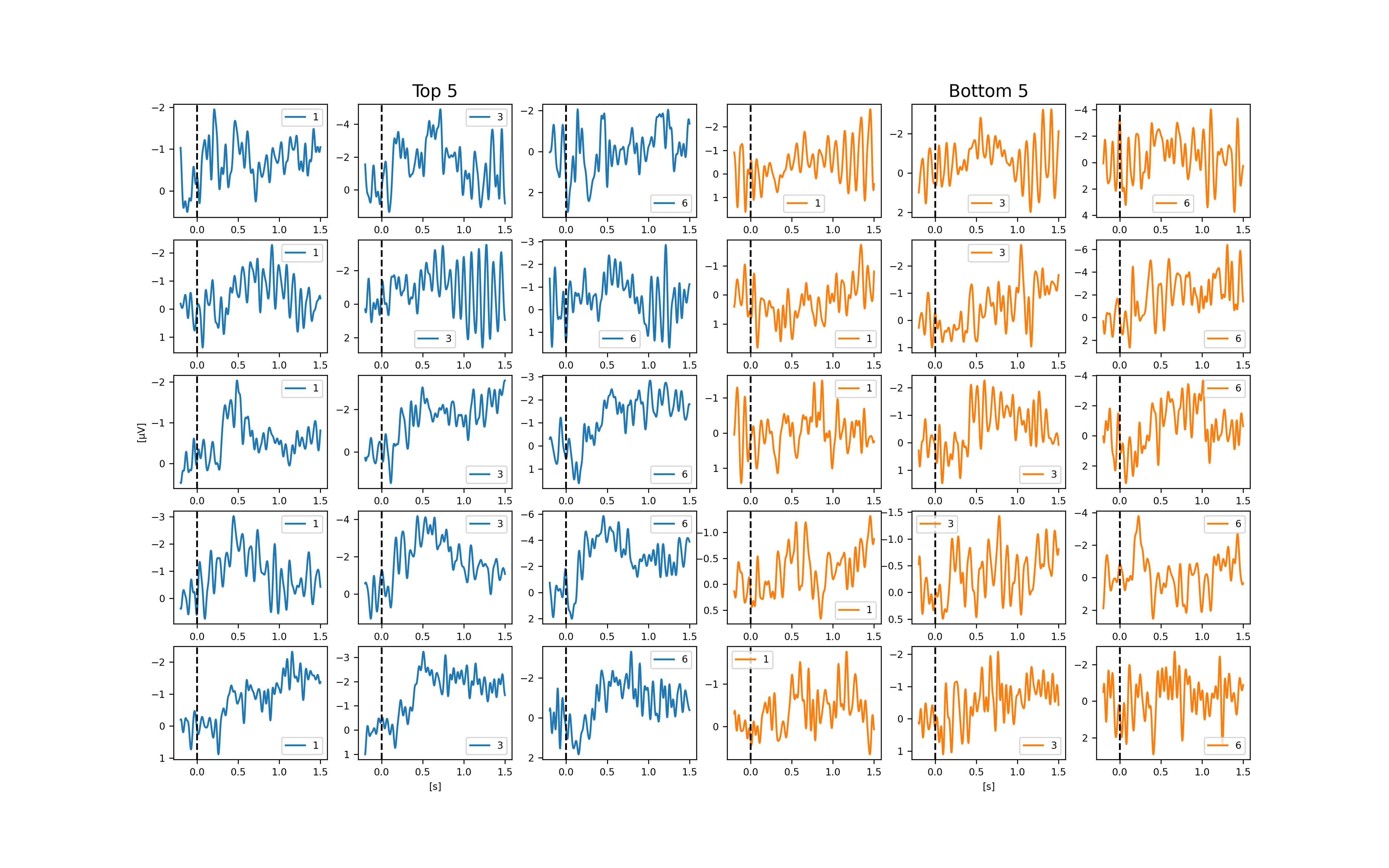}
    \caption{Top 5 and Bottom 5 from Adam, 2018}
    \label{Adam-2018-Top5Low5}
\end{figure}

Given the weak but consistent correlation between CDA amplitude and individual performance, we looked at the individual level to explore the variability on CDA across participants. On figures \ref{Adam-2018-Top5Low5}, \ref{Villena-2019-Top5Low5}, \ref{Balaban-2019-Top5Low5}, and \ref{Feldmann-2020-Top5Low5} in supplementary material we did plot the CDA amplitude of the \textit{top 5} and \textit{bottom 5} participants of four different studies reviewed here (Adam, 2018; Villena-Gonzalez, 2019; Balaban, 2019; Feldmann-Wüstefeld, 2020), including three different conditions for each. These figures are exploratory and no further statistics were performed, however, we believe it gives more perspective on the CDA shape and its variability across subjects and across studies. The blue graphs on left represents the CDA of the 5 participants with the best performance on the task (top 5) and the orange graphs represents the CDA for the 5 participants with the worst performance (bottom 5). The top left graphs in blue on the first row represent the best participant, performance wise, and the bottom right graphs in orange the last row represent the participant with the lowest performance score. By visually looking at the graphs, it would be difficult to decipher a clear trend of CDA amplitude on performance, explaining the weak correlation.
It is important to note here that the y axis (CDA amplitude in microvolt) was not fixed since the peak-to-peak amplitude varies quite significantly between participants and finding a once-size-fits-all range ends up hiding the shape of the CDA, which is what we seek to showcase here. We invite the reader to pay a close attention to the value on the y axis before drawing any conclusion. 


\section{Discussion} 
\label{sec:discussion}

As shown on previous figures, all the reproduced studies showed a clear CDA across different VWM paradigms. Given the prevalence of CDA decay during retention and re-increase of CDA amplitude during recall across studies, we suggest that the CDA could have more to do with \textit{updating} (or refreshing) the mental representation than actually \textit{maintaining} it. For example, in Balaban, 2019, when the items are separated, the CDA increases and reaches a higher peak than for integrated shape as expected if the CDA amplitude correlates with the number of items in working memory. However, by the end of the retention phase, just before being probed, the two conditions have pretty much the same CDA amplitude, which would normally indicate that at that point in time there is the same number of items in working memory, which is not the case. Moreover, in Feldmann-Wüsterfeld 2018 Experiment 1, we see a cumulative effect of the CDA amplitude each time an \textit{update} occurs, whether from external stimuli or internally induced for recall. The CDA decay we observe consistently across studies (also other CDA studies not included here) seems coherent with WM models where oscillatory processes are keeping the information 'alive' via cell assembly firing together synchronizing lower frequencies in the theta and alpha bands (4 Hz to 12 Hz) with higher frequencies ($~$40 Hz). The cell assembly passively decays, requiring regular updates before it decays too much and the information is lost (\cite{lisman1995storage, luck2013visual}).

Interestingly, the CDA amplitude during recall seems to have the same amplitude for all conditions which is not the case during the retention phase (see Figures \ref{recall-Balaban2019}, \ref{recall-Hakim2019}, \ref{recall-Feldmann2020}, and \ref{recall-Feldmann2018} top-left corner). We suggest that during the initial identification of the items, the indexing might happen one by one in a serial fashion, increasing the CDA in a cumulative way for each item being indexed, explaining the different amplitudes for different number of items. During recall, the update or refreshing of working memory is internally induced and could bring back all the items at once, hence showing the same CDA amplitude for all conditions. Moreover, the rate of ascend of the CDA is significantly higher during recall, as if the items were brought back in a more parallel fashion to memory. Without further investigation, it is hard to say if this higher rate of ascend (i.e. slope) is more due to a biological factor, like the neurons already being primed and active or if it is more of a cognitive factor, like recalling all the items at once bypassing a lot of neural processes present during the identification part.

As mentioned earlier, this sample of studies does not represent an exhaustive list of CDA studies with publicly available EEG datasets and this review could suffer from a bias given that many of the authors of selected studies are present and/or previous colleagues and collaborators. 



There are several barriers for researchers to make their research reproducible, however, one that can be alleviated is the lack of knowledge and awareness about the available tools and best practices for reproducibility. In this study, we thought we would share what others have done and how they've done it, with the hope of helping other fellow researchers to do the same. Hopefully this study will reduce the fear of the extra steps required to make your study reproducible as well as highlight the value of taking these extra steps enabling one's work to generate a greater impact on the field.

Moreover, the FAIR principles (\cite{wilkinson2016fair}) and the Brain Imaging Data Structure (BIDS) (\cite{gorgolewski2016brain}) both provide guidelines and standards on how to acquire, organize and share brain-related data and code. As the amount of recorded brain data keep increasing around the world and becomes more openly accessible it is important to have best practices to reduce the friction and wasted time. In addition to reproducibility aiming at validating or invalidating scientific evidences, data mining leveraging new approaches such as artificial intelligence (AI) and deep learning (DL) can benefit from having access well documented and openly accessible brain datasets (\cite{roy2019deep}). Trainees are also benefiting tremendously from reproducible experiments. Unintuitively however, most young trainees venturing in a new field think that it will be quick and easy to reproduce someone else's work and then take it from there to either modify or improve it. Unfortunately, they quickly realize that despite being theoretically easy, it isn't. There are always a multitude of complications from different operating system (OS) compatibility issues to versions of software libraries to missing parts of the code or lack of documentation making it hard to understand the code and its logic. A better reproducibility culture and best practices can help reduce significantly such friction and waste of time. 

\textbf{Reproducibility recommendations.} (1) EEG preprocessing; in most published papers, the preprocessing steps are well described, however, many studies have a "visual inspection" part to remove some of the contaminated data. This subjective step can make reproducing the results difficult and we encourage researchers to also share the preprocessed EEG files of each participants in order to be able to compare where the results start diverging when failing to reproduce the results. This obviously increases the size of data being stored and shared but adds a lot of value. (2) Excluded subjects; in their documentation file (e.g. README.md), researchers should make clear which subjects have been excluded from the analysis. Usually, in the published paper there is a mention like "3 subjects were removed because of ...", while this informs the reader about the number of subjects that were included, if the raw data of these subjects are included in the data folder, it puts the burden on someone trying to reproduce the study to find the excluded ones. We've encountered four ways to dealing with this issue while reproducing the studies for this review. (a) Not sharing the raw data of excluded subjects. In most cases we would not recommend this as there might be useful parts in the data. It really depends on the reason for excluding the participants from the study. (b) Making a specific \textit{Excluded} folder and inserting the excluded subjects' data in it. (c) Identifying the excluded participant in the documentation file. Sometimes that information is available somewhere in the code as a \textit{if} statement like \textit{if fname == "participant\_X"}, however this information should be brought forward and featured in the main documentation file. (d) Identifying the excluded participants in the published paper. This works well when one or two participants are excluded with a mention like "participant \#12 was excluded because of ..." but doesn't scale well for multiple participants. (3) Document triggers/events; the triggers/events allowing to epoch the data properly should be clearly identified. Such information is usually a mix of hardware triggers from the EEG file and information from a behavioural file (e.g. csv file). For example, in most studies the side and the set size of each trial is available in the EEG file (saved as a signal with the EEG hardware) but the performance of the trial is available in a behavioural file. Such structure should be clearly described the main documentation file. (4) Performance scoring code; there are many ways to assess performance (e.g. K Score) and it would be beneficial to share the corresponding code to avoid confusion. In most studies, the formula is described in the paper but the corresponding code is not available. (5) Online repository platform; we recommend using OSF to share the data. \textit{OSF is a free, open platform to support your research and enable collaboration}, as stated on their website. All of the reviewed studies, but one, used OSF to share both the data and the code. Villena-Gonzalez used Mendeley Data instead. (6) README file; we recommend having a README.md or .pdf file in the root folder explaining the crucial information about the data with reproducibility in mind. We provide as good example from Adam 2018 below. We took a print screen of the beginning of the file. Their README file provides a clear explanation of the files and the data contained in these files saving precious time of investigating to figure it out. Beware of the copy/paste across experiments! Their first sentence shows a good example of a copy/paste that wasn't edited. In this case, the information isn't misleading and the mistake is pretty obvious but in other contexts it can be very counterproductive. (7) Raw data; researchers should include the raw data from the EEG recording device and not an exported version from EEGLab or else. Even if the data hasn't been modified and has been exported as is, it creates doubt and questions for no added value. The preprocessed data will be exported from tools like EEGLab or .mat files, however the raw data files should be the one coming from the recording device directly. 


Finally, our objective with this review, aside from better understanding the cognitive mechanisms at play during a VWM task and the key role of CDA, was to evaluate the CDA reliability as a potential candidate for brain-computer interfaces (BCIs). Many other ERPs and neural correlates are being used in brain-computer interface paradigms, however, to our knowledge, the CDA has never been used in a BCI context. We have, however, seen a recent interest in using machine learning for CDA classification with regards to the number of items held in WM (e.g. \cite{adam2020multivariate}. We believe that within certain contexts, CDA could be used to enhance passive BCIs (\cite{zander2011towards}).


\section{Conclusion} 
\label{sec:conclusion}
With this study we were able to look at CDA across different VWM tasks, from different groups of subjects, recorded by different groups of researchers using different EEG equipment. By using the same simple independent pipeline for all studies, we have shown that CDA is a robust neural correlate of visual working memory. As for its exact role and meaning, more research is required. 

Moreover, having access to all these datasets allowed us to look beyond the usual numerical CDA mean amplitude over a window of interest but to also observe the decay happening shortly after the CDA has peaked while the participant still has to maintain the information. Even more importantly, it has also allowed us to look at the recall phase, which none of the reviewed studies had done (or mentioned). We found that a similarly shaped CDA is also present and in some cases of higher amplitude than initial CDA during the identification and tracking or retention phase. It was a surprise to us that none of the reviewed studies discussed this phenomena, especially given the size of the effect.

Finally, all the code and figures generated for this review is available online on our repository. Many more figures that we did not include in the manuscript to keep it as short and concise as possible for better readability, are available on the repository. 

Code and analysis: \textit{https://github.com/royyannick/cda-reprod}


\section{Funding}
This work was supported by the Natural Sciences and Engineering Research Council of Canada (NSERC-RDC) (reference number: RDPJ 514052-17) and an NSERC Discovery fund.

\section{Conflict of Interest}
The authors declare that there is no conflict of interest.

\bibliography{main}

\section{Supplementary Material}


\begin{figure} [!ht]
    \centering
    \includegraphics[width=\textwidth]{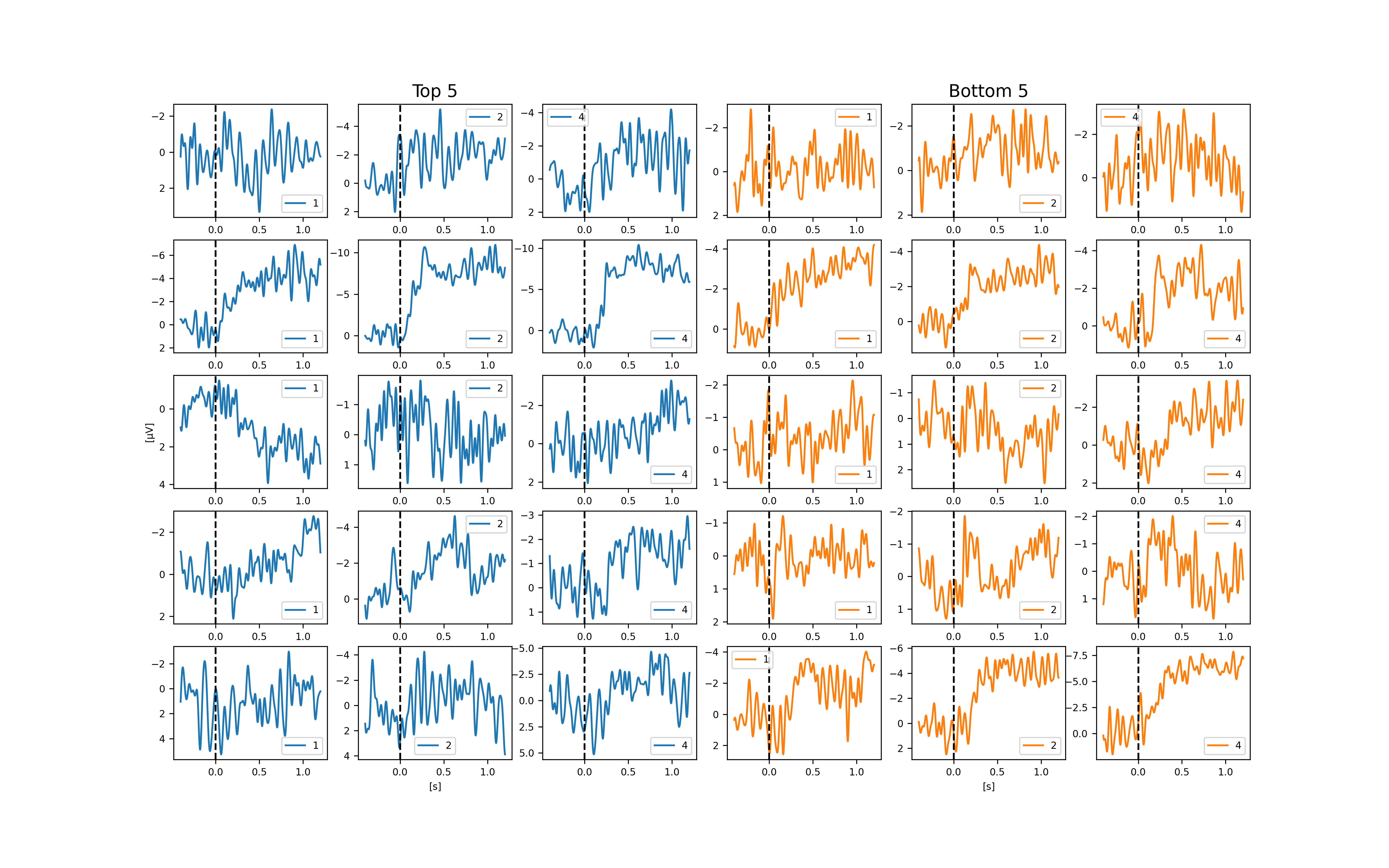}
    \caption{Top 5 and Bottom 5 from Villena-Gonzalez, 2019}
    \label{Villena-2019-Top5Low5}
\end{figure}

\begin{figure} [!ht]
    \centering
    \includegraphics[width=\linewidth]{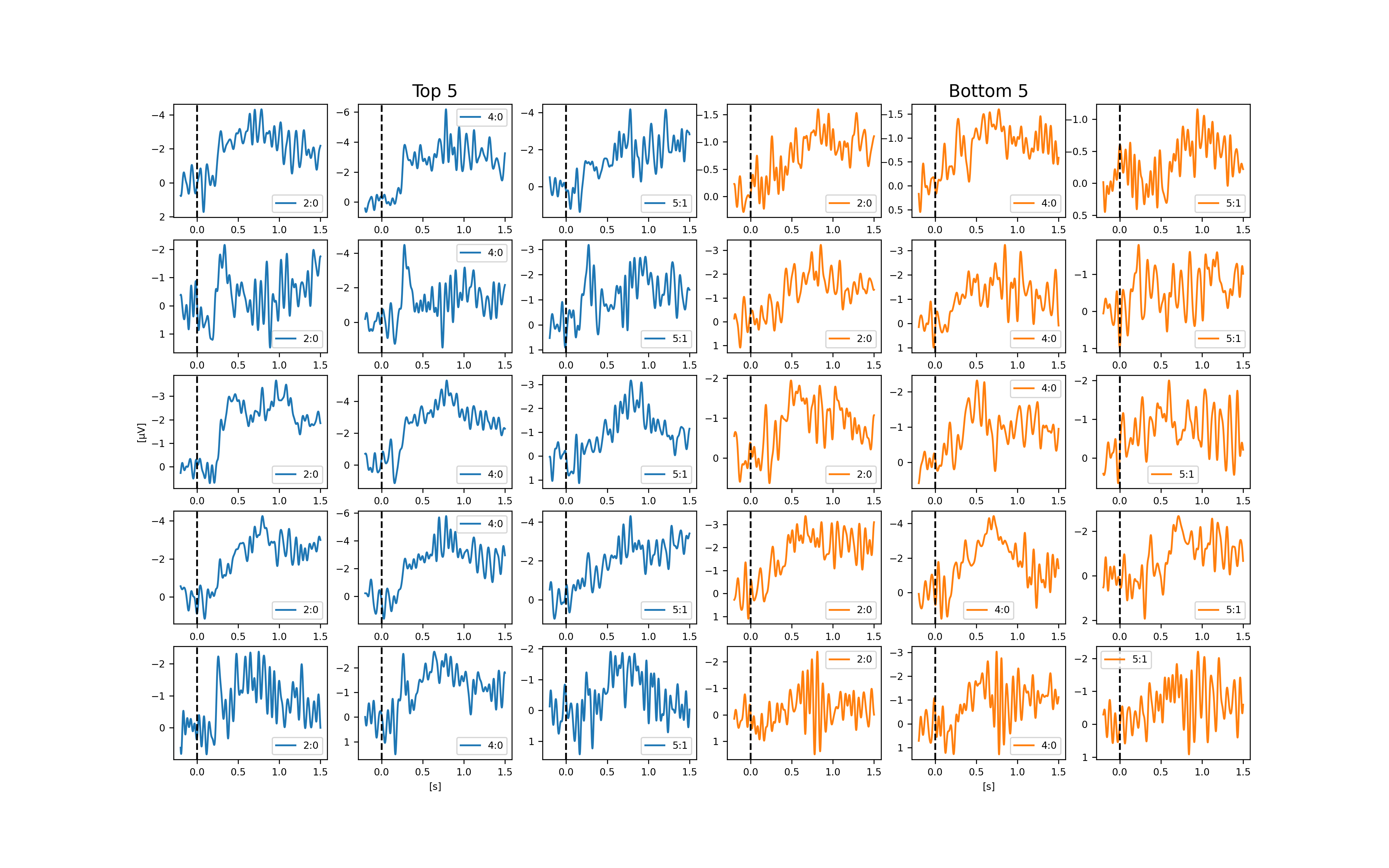}
    \caption{Top 5 and Bottom 5 from Feldmann-Wusterfel 2020}
    \label{Feldmann-2020-Top5Low5}
\end{figure}

\begin{figure} [!ht]
    \centering
    \includegraphics[width=\linewidth]{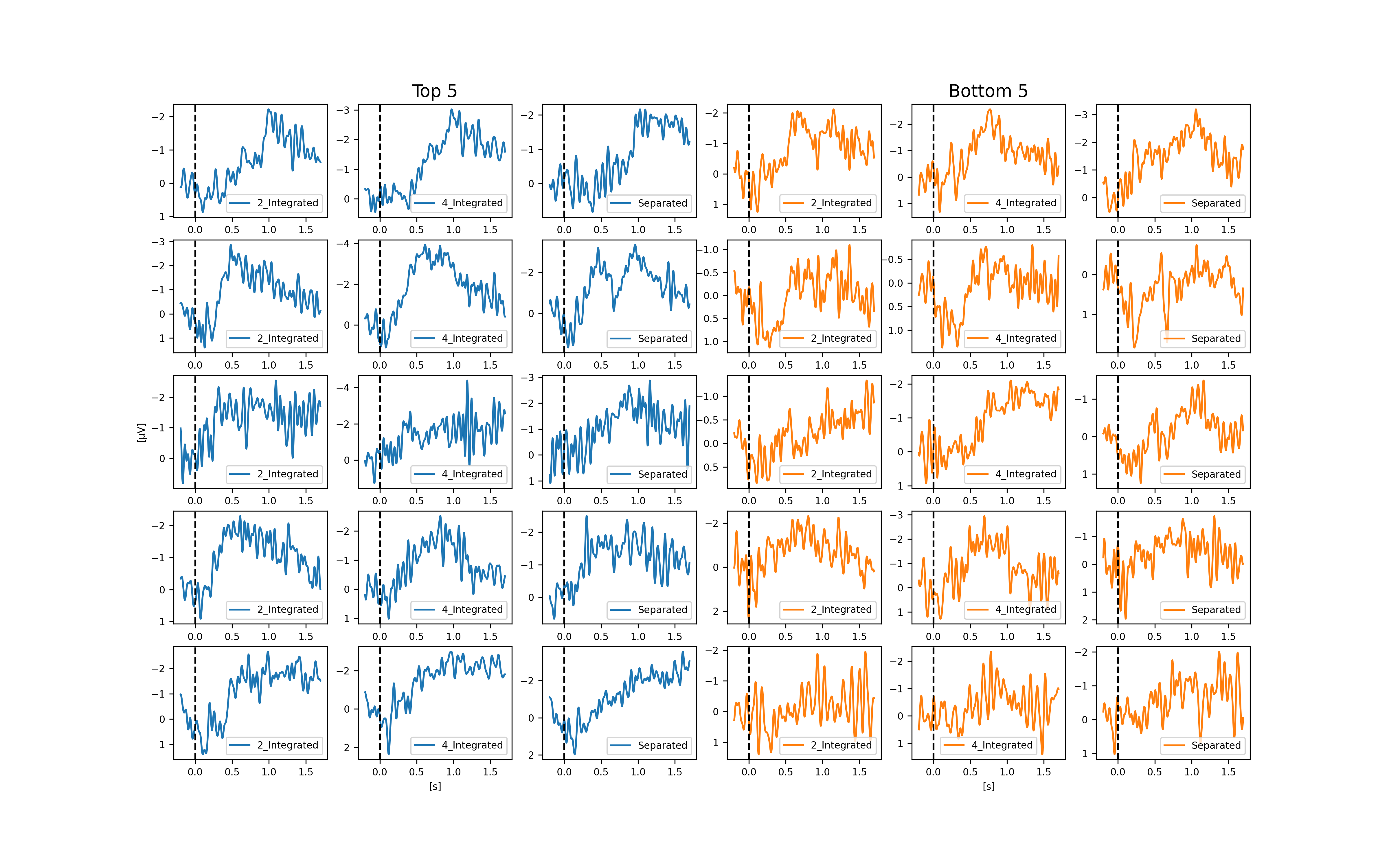}
    \caption{Top 5 and Bottom 5 from Balaban 2019 Exp. 2}
    \label{Balaban-2019-Top5Low5}
\end{figure}


\begin{figure} [!htb]
    \centering
    \includegraphics[width=0.8\textwidth]{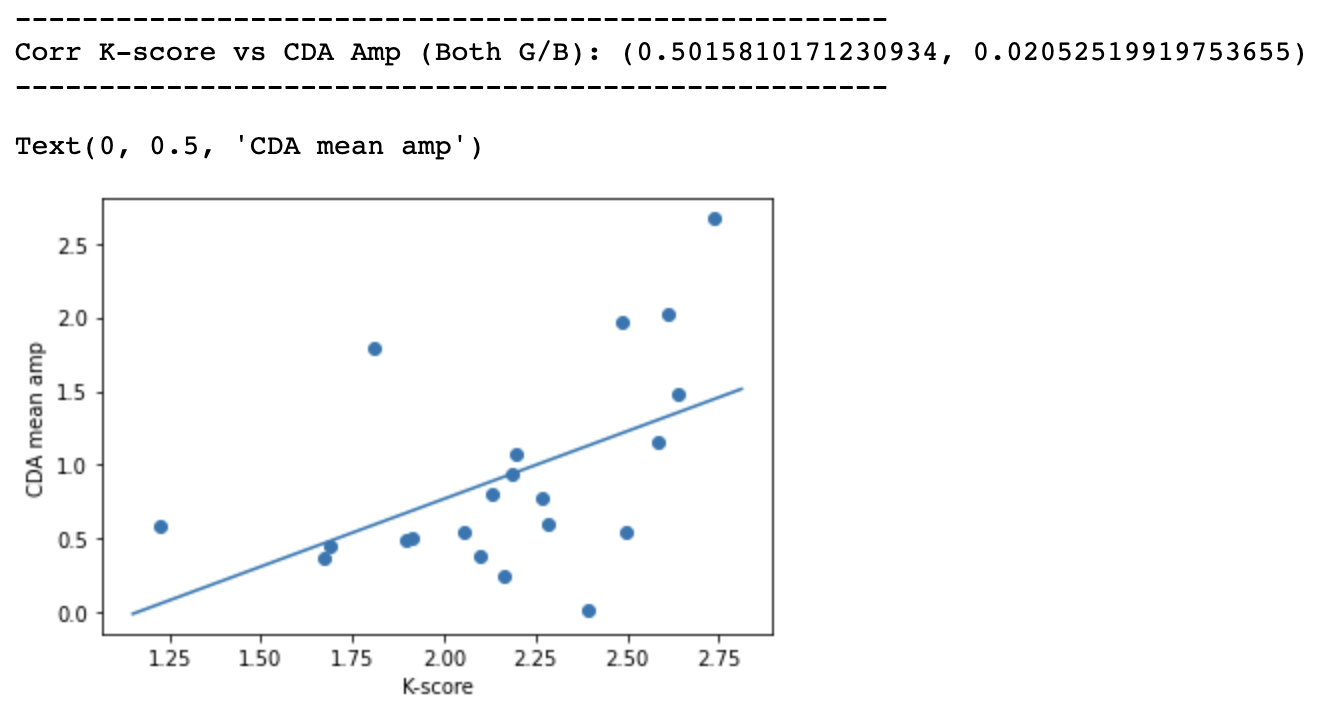}
    \caption{CDA Amplitude vs Performance - Feldmann-Wüstefeld, 2020}
    \label{Feldmann-2020-stats}
\end{figure}
\begin{figure} [!htb]
    \centering
    \includegraphics[width=0.8\textwidth]{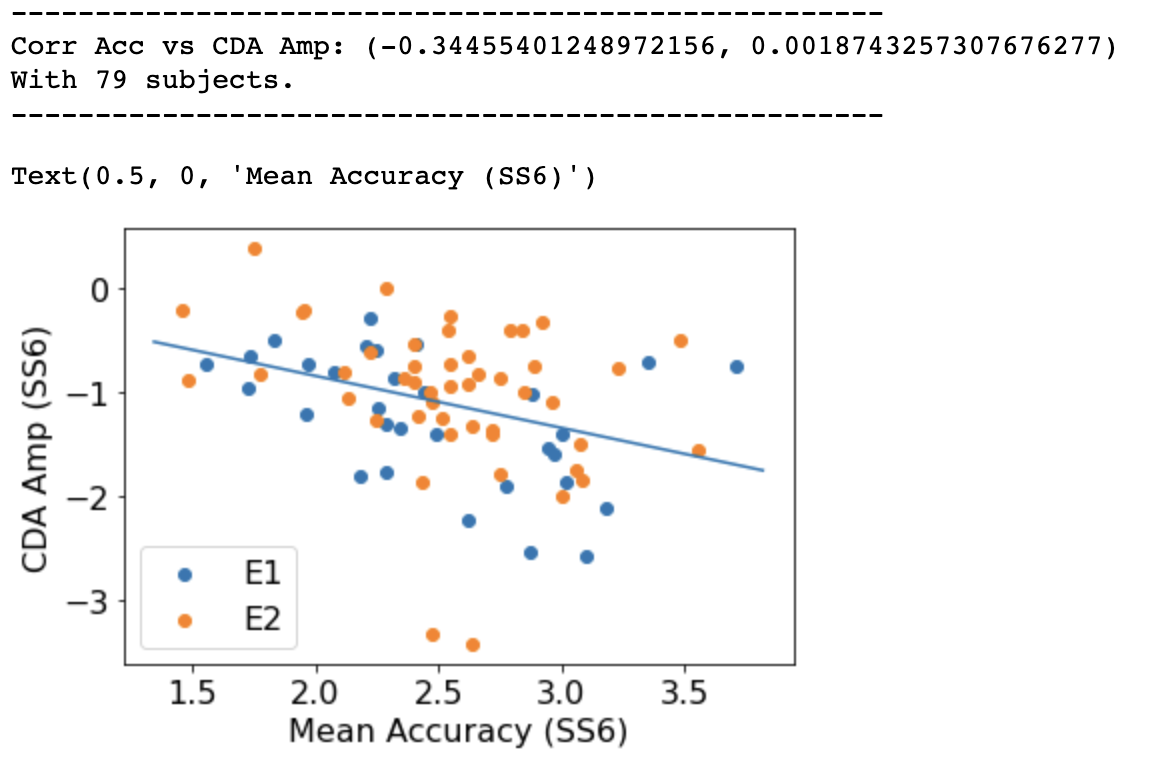}
    \caption{CDA Amplitude vs Performance - Adam, 2018}
    \label{Adam-2018-stats}
\end{figure}


\begin{figure} [!htb]
    \centering
    \includegraphics[width=\textwidth]{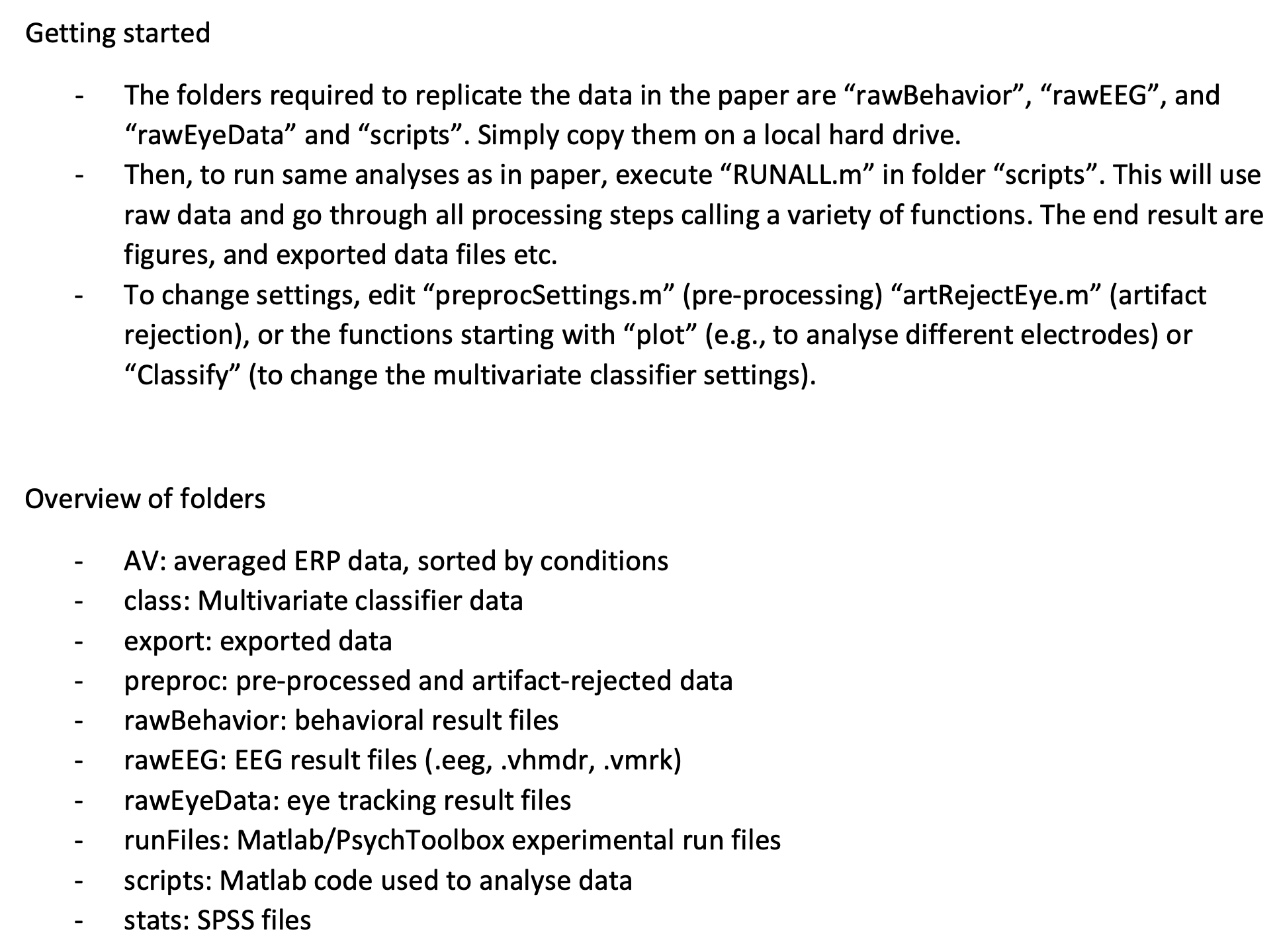}
    \caption{Getting Started from Feldmann-Wüstefeld, 2020}
    \label{FW2020-ReadMe}
\end{figure}

\begin{figure} [!htb]
    \centering
    \includegraphics[width=\textwidth]{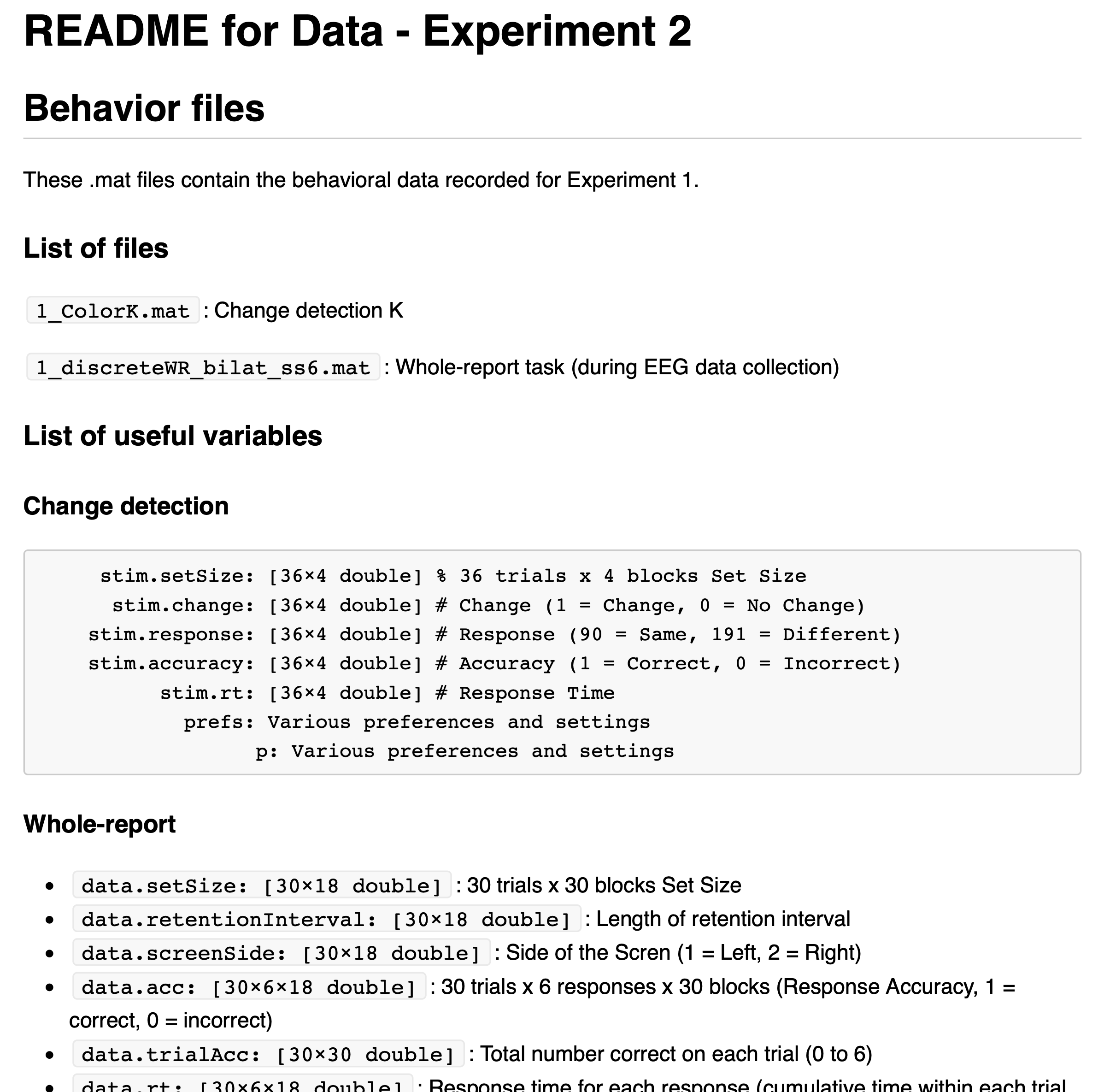}
    \caption{README from Adam, 2018}
    \label{A2018-ReadMe}
\end{figure}

\end{document}